%% file: main.tex
\documentclass{article}
\usepackage{arxiv}
\usepackage[backend=biber, style=numeric, maxbibnames=1,url=true]{biblatex}
\usepackage{hyperref}

\usepackage{times}
\usepackage{fullpage}

\usepackage{pifont}
\usepackage{url,tikz}
\usepackage{graphicx}
\usepackage{booktabs}
\usepackage{amsfonts}       
\usepackage{nicefrac}      
\usepackage{xcolor}         
\usepackage{graphicx, amsfonts, amsmath,amsthm, amssymb} 
\newcommand\numberthis{\addtocounter{equation}{1}\tag{\theequation}}
\usepackage{xcolor}
\usepackage{algorithm}
\usepackage[noend]{algpseudocode}
\usepackage{mathrsfs, comment,csquotes}
\usepackage{framed}
\usepackage{varwidth}
\usepackage{nccmath, mathtools}
\usepackage{subcaption}

\usepackage{multirow, multicol}
\input{defs}

\newcommand{\david}[1]{\textcolor{blue}{David: }\textcolor{blue}{#1}}

\newcommand{\cmark}{\textcolor{green}{\ding{51}}} 
\newcommand{\xmark}{\textcolor{red}{\ding{55}}}   

\usepackage{tikz}
\usepackage{amsmath}
\usepackage{tcolorbox}

\begin{document}

\date{}

\title{\Large \bf \name: Sanitizing Sensitive Prompts for LLMs}

\author{
Amrita Roy Chowdhury\thanks{These authors have contributed equally to this work.} \\ University of Michigan, Ann Arbor \\
\And
David Glukhov$^{*}$ \\University of Toronto and Vector Institute \\
\And
Divyam Anshumaan$^{*}$ \\University of Wisconsin-Madison \\
\And
Prasad Chalasani \\Langroid Incorporated \\
\And
Nicolas Papernot \\University of Toronto and Vector Institute\\
\And
Somesh Jha \\University of Wisconsin-Madison \\
\And
Mihir Bellare \\University of California, San Diego
}


\maketitle

\begin{abstract}
The rise of large language models (LLMs) has introduced new privacy challenges, particularly during \textit{inference} where sensitive information in prompts may be exposed to proprietary LLM APIs. In this paper, we address the problem of  formally protecting the sensitive information contained in a prompt  while maintaining response quality. To this end, first, we introduce a cryptographically inspired notion of a \textit{prompt sanitizer} which
transforms an input prompt to protect its sensitive tokens. Second, we propose \name, a novel system that implements a prompt sanitizer, focusing on the sensitive information
that can be derived solely from the individual tokens.
\name~ categorizes sensitive tokens into two types: (1) those where the LLM's response depends solely on the format (such as SSNs, credit card numbers), for which we use format-preserving encryption (\FPE); and (2) those where the response depends on specific values, (such as age, salary) for which we apply metric differential privacy (mDP). Our evaluation demonstrates that \name~is a practical method to achieve meaningful privacy guarantees, while maintaining high utility compared to unsanitized prompts, and outperforming prior methods.

 
\end{abstract}

\input{NDSS_revision/1_Introduction}
\input{NDSS_revision/2_Background}

\input{NDSS_revision/3_ProposedScheme}
\input{NDSS_revision/4_Preempt}

\input{NDSS_revision/4_Security_Utility}
\input{NDSS_revision/4_Evaluation_Short}

\input{NDSS_revision/5_Discussion}
\input{NDSS_revision/7_Conclusion}


\section*{Availability}

An implementation of \name~and experimental results can be found at~\cite{Preempt}.

\section*{Acknowledgements}
David Glukhov and Nicolas Papernot would like to acknowledge their sponsors, who support their research with financial and in-kind contributions: Amazon,
Apple, CIFAR through the Canada CIFAR AI Chair, Microsoft, Meta, NSERC through the Discovery Grant and two Alliance Grants with ServiceNow and DRDC and with CSE, the Ontario Early Researcher Award, and the Schmidt Sciences foundation through the AI2050 Early Career Fellow program. Resources used in preparing this research were provided, in part,
by the Province of Ontario, the Government of Canada through CIFAR, and companies sponsoring the Vector Institute.

Divyam Anshumaan and Somesh Jha are partially supported by DARPA under agreement number 885000, NSF CCF-FMiTF-1836978 and ONR N00014-21-1-2492.

\section{Ethical Considerations}
We provide a framework for enabling users to sanitize sensitive tokens before submitting prompts to proprietary LLMs. To evaluate our framework, we generated synthetic data  for sensitive tokens, avoiding issues of using or releasing sensitive data of any individuals. While theoretically sound, there still exist practical limitations of our sanitization procedure, for example imperfect detection of types by NER methods, that could result in sensitive attributes being leaked. Furthermore, LLMs may still behave more erratically on tail events and it is possible that sanitization could result in inputs that significantly alter model behavior. If these limitations are ignored, usage of our framework could lead to a false sense of safety by the user. As such, we do not make claims that our method is always effective and we provide empirical evidence demonstrating failure modes. Furthermore, our theoretical analysis provides a useful step forward for the problem of prompt sanitization, and we highlight several often overlooked issues of prior work on text sanitization work. 

We also emphasize that our toy Financial settings is not intended to illustrate recommended use cases for our approach, particularly as questions regarding financial and medical decision making should be addressed to certified professionals as opposed to LLMs. Nevertheless, many users may still decide to interact with LLMs in this manner and risk revealing sensitive information in the process. We studied these settings specifically for the purpose of understanding the how sanitiztion affects model decision making in such ill-specified settings where there is no ``correct answer'' but a model must still make a decision. As such, our results reflect reasoning and robustness of LLMs to sanitization methods.
\printbibliography

\input{NDSS_revision/Appendix_ndss_revision}
\end{document}

%% file: defs.tex
\usepackage{xcolor,comment}
\usepackage{resizegather}
\newtheorem{defn}{Definition}
 
\newtheorem{theorem}{Theorem}

\usepackage{subcaption}
\def \LDP{LDP}
\def \DP{DP}

\def \FPE{\textsf{FPE}}

\def \T{\textsf{T}}
\def \m{f}
\def \t{\tau}

\def \Set{\textsf{S}}
\def \s{\sigma}
\def \ss{\boldsymbol{\sigma}}

\def \RdS{\textsf{RdS}}
\def \p{\boldsymbol{\rho}}
\def \r{r}
\def \PS{\textsf{PS}}
\def \FPEs{\textsf{FPEs}}
\def \LLM{\text{LLM}}
\def \AT{\mathcal{M}_\tau}

\def \M{\mathcal{M}}

\def \Mpre{\mathcal{M}_{\textsf{Pre}}}
\def \Mpost{\mathcal{M}_{\textsf{Post}}}
\def \Mdp{\mathcal{M}_{\epsilon}}
\def \U{U}
\def \Q{\mathcal{Q}}
\def \mLDP{mLDP}

\def \Nf{\mathcal{N}}
\def \N{\textsf{N}}
\def \RtE{\textsf{RtE}}
\def \CW{\textsf{CW}}
\def \name{Pr$\epsilon\epsilon$mpt}
\def \MP{\textsf{MP}}
\def \MR{\textsf{MR}}
\def \SPI{\textsf{SPI}}
\def \PRP{\textsf{PRP}}
\def \advppt{\mathcal{A}}
\def \K{\textsf{K}}
\def \Gen{\textsf{G}}
\def \V{\textsf{V}}
\def \A{\mathcal{A}}

\def \L{\mathcal{L}}
\def \r{\boldsymbol{\upsilon}}
\def \NS{\perp}

\def \DF{\textsf{D}_{\mathcal{F}}}
\def \EF{\textsf{E}_{\mathcal{F}}}
\def \GenF{\textsf{G}_{\mathcal{F}}}

\newcommand{\E}{\textsf{E}}
\newcommand{\D}{\textsf{D}}
\newcommand{\todo}[1]{\textcolor{blue}{#1}}
\newcommand{\arc}[1]{\textcolor{blue}{ARC: #1}}

\newcommand{\Initialize}{\textsc{Initialize}}
\newcommand{\Finalize}{\textsc{Finalize}}
\newcommand{\Sanitize}{\textsc{Sanitize}}

\newcommand{\gamePS}[1]{\mathbf{G}^{\mathrm{pp}}_{#1}}

\def\bits{\{0,1\}}

\newcommand{\genAdv}[3]{\mathbf{Adv}^{\mathrm{#1}}_{#2}(#3)}
\newcommand{\ppAdv}[2]{\genAdv{pp}{#1}{#2}}

\newcommand{\squishlist}{
	\begin{list}{$\bullet$}
		{
			\setlength{\itemsep}{1pt}
			\setlength{\parsep}{2pt}
			\setlength{\topsep}{4pt}
			\setlength{\partopsep}{0pt}
			\setlength{\leftmargin}{1.5em}
			\setlength{\labelwidth}{1em}
			\setlength{\labelsep}{0.5em} } }
	
\newcommand{\squishend}{
\end{list}  }

%% file: NDSS_revision/1_Introduction.tex
\section{Introduction} 
\label{sec:intro}
The recent advent of large language models (\LLM s) have brought forth a fresh set of challenges for protecting users' data privacy. \LLM s and their APIs present significant privacy concerns at \textit{inference time}, which are fundamentally distinct from the well-documented risks of training data memorization~\cite{carlini2021extracting,Mem1,lee2021deduplicating,mem2}. While the potential adversary in training data scenarios could be any API user, the threat during inference primarily stems from the model owner—typically the organization hosting the \LLM. This inference stage poses a significant privacy risk, as \textit{prompts} in natural language may include various types of sensitive information, from personally identifiable data like SSNs or credit card numbers to personal health or financial details. 

The ensuing privacy threat is exacerbated with the growing use of in-context learning, that involves presenting the \LLM~with a few training examples as part of the prompt during inference~\cite{brown2020language}. This has shifted some of the concerns around privacy of training data from training time to inference time.  Furthermore,  the consumer-facing nature and widespread accessibility~\cite{cui2023chatlaw,10.1093/eurjcn/zvad022,kamalov2023new,singhal2023towards,jeblick2023chatgpt} of \LLM s have significantly amplified the scope of these privacy risks. 
What renders the privacy risks particularly potent is the general lack of awareness among users, leading to unwitting disclosure of sensitive information~\cite{barrett2023identifying}. Consequently, certain countries, such as Italy~\cite{Italy}, along with financial institutions ~\cite{jpmorgan-ban, goldman-ban}, government agencies~\cite{ssa-ban, doe-ban, ussf-ban}, medical institutions ~\cite{Ausban} as well as companies, such as Samsung~\cite{Samsung,Samsung2}, Amazon~\cite{Amazon} and Apple~\cite{apple-ban}, have prohibited the use of proprietary LLMs altogether, underscoring the significance of these privacy concerns.
\begin{figure}[t]
    \centering
    \includegraphics[width=0.7\linewidth]{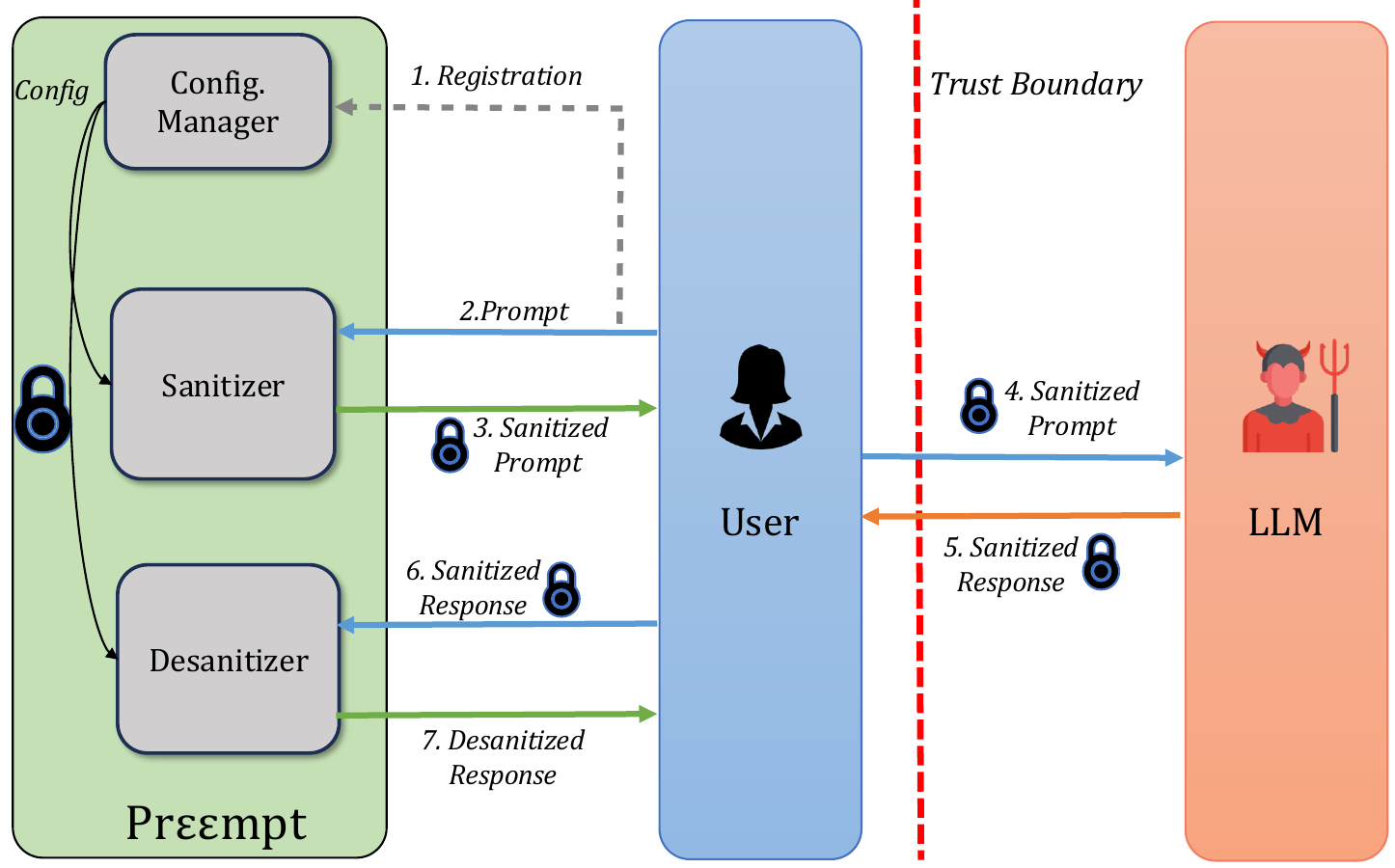}
    \caption{Overview of \name: Users begin with a one-time registration to set up configurations, which are used in all subsequent interactions. Users can then submit prompts to \name~and receive their sanitized versions, which are safe to be provided to the untrusted \LLM. The LLM's responses (to the sanitized prompts) can then be desanitized to recover high-utility responses. }
    \label{fig:overview} \vspace{-0.5cm}
\end{figure}

Prior work on privacy-preserving LLMs~\cite{anil2021largescale,li2022large,Hoory2021LearningAE,ramaswamy2020training,shi2022selective,mcmahan2018learning,yu2022differentially}, has primarily focused on mechanisms during training. Unfortunately, these training-time mechanisms can only protect the (pre)-training data: data provided as part of the prompt poses additional privacy risks that are not addressed by these mechanisms~\cite{Duan2024}. While recent research has begun to address the privacy of prompts, solutions based on homomorphic encryption and secure multi-party computation~\cite{hao2022iron,chen2022x,cryptoeprint:2023/1147} are computationally expensive in practice, with state-of-the-art techniques taking over {\it sixteen} minutes for a single inference on BERT~\cite{Bolt}. More efficient solutions either lack formal privacy guarantees~\cite{lin2024emojicryptpromptencryptionsecure, zhang2024latticegencooperativeframeworkhides}, require changes to current LLM APIs~\cite{mai2023split}   or make impractical design choices~\cite{shen2024thiefkeeperbalancingusability, kan2023protecting,chen2023hide} (see Sec. \ref{sec:relatedwork} for more details).

 To this end, we make the following contributions. First, we introduce a cryptographically inspired notion of  a \textit{prompt sanitizer} that takes a prompt and transforms it in a way that protects \textit{sensitive tokens} yet still preserves the ability of the LLM to make a useful prediction. We provide a formal analysis of both its privacy and utility. Second, we propose \name\footnote{\textbf{P}rivacy-P\textbf{re}serving Pro\textbf{mpt}}, a system that instantiates a prompt sanitizer. 
 To the best of our knowledge, Pr$\epsilon \epsilon$mpt is the \textit{first} prompt sanitizer with formal privacy guarantees. As the first step in this direction, we focus on the sensitive information that can derived \textit{solely} from the individual tokens. It is important to note that addressing this aspect is paramount as it poses the most \textit{immediate risk} and represents a ``low-hanging fruit'' for potential adversaries. This is because an adversary can exploit the sensitive tokens (such as SSN, credit card number) independently, without needing to process or access additional context from the prompts. Moreover, in many settings, sensitive information is often restricted to structured (e.g., tabular) data that can be extracted as tokens—for instance, in financial Q/A tasks as evaluated in our experiments. The task of handling privacy risks stemming from the contextual linguistic semantics\footnote{This refers to cases where individual tokens may not be sensitive but, when considered in the context of the full prompt, could leak information because of the underlying natural language semantics.} of the entire prompt~\cite{mireshghallah2023llms,brown2022does} is left as future work.
 
 \name~operates on the assumption that sensitive tokens can be categorized into two types: $(1)$ tokens for which the \LLM's response depends \textit{solely} on their format. (e.g., SSN, credit card number), $(2)$ tokens where LLM's response depends on the specific numerical value itself (e.g., age, salary). Consequently, we propose encrypting the former using format-preserving encryption~\cite{BRRS2009}: a type of property-preserving encryption scheme where the ciphertext and the plaintext have the same format. For example, the ciphertext of a 16-digit credit card number encrypted under a \FPE~scheme would also be a 16-digit number. 
Tokens of the second type are sanitized using differential privacy (DP)~\cite{Dwork}, which is the state-of-the-art technique for achieving data privacy. Specifically, we employ a relaxation of DP, called metric DP~\cite{chatzikokolakis2013broadening}. Metric DP protects pairs of inputs that are ``similar'' based on a distance metric, meaning that the sanitized token will remain similar to the original token. This approach maintains the relevance of the responses generated to the original prompt while providing meaningful privacy guarantees. 

We demonstrate the practicality of \name~through empirical evaluation. Specifically, we evaluate four types of tasks: translation, retrieval  augmented generation (RAG), long-context reading comprehension Q/A and multi-turn financial Q/A. We observe that \name's sanitization mechanism preserves the utility of responses across all tasks. For instance, the BLEU scores~\cite{papineni-etal-2002-bleu} for sanitized prompts are nearly identical compared to baseline unsanitized prompts for a German language translation task with GPT-4o.~
When prompted with \name~sanitized prompts, all RAG tasks achieved 100\% accuracy. \name~is also quite successful in long-context and multi-turn conversation tasks. For example, responses based on \name~ processed reference texts used in long-context Q/A has a similarity score of $0.934$ compared to responses based on unsanitized text, outperforming a contemporary method~\cite{siyan2024papillon} (PAPILLON) without any additional overheads. 
\\

%% file: NDSS_revision/2_Background.tex
\vspace{-0.3cm}
\section{Background}
\label{sec:background}
\noindent\textbf{Notation.}
Let $\V$ be the vocabulary (tokens) of a language model and $\V^*$ the set of possible strings over $\V$ (recall that a prompt and its response are strings over $\V$).  We represent a sequence of tokens $\ss \in \V^*$ with a boldface. Let $f$ be a \LLM~and $\p \in \V^*$ be a prompt for it. A prompt is a sequence of tokens from $\V$, i.e., $\p=\langle \sigma_1, \cdots, \sigma_n\rangle, \sigma_i \in \V, \forall i \in [n]$. Let $\mathbb{P}(\V)$ denote the space of all probability distribution over $\V$. 

\subsection{Language Model}
\label{sec:LLM}
\begin{defn} A language model $\m$ is an auto-regressive model over a vocabulary $\V$. It is a deterministic algorithm that takes a prompt $\p \in \V^*$ and tokens previously produced by the model $\ss \in \V^*$ as input, and outputs a probability distribution $p = \m(\p,\ss)$ for $p \in \mathbb{P}(V)$.
\end{defn}

A language model's response to a prompt $\p$ is a random variable $\ss \in \V^*$ that is defined algorithmically as follows. We begin with an empty sequence of tokens $\ss = \langle \rangle$. As long as the last token in $\ss$ is not $\bot$ (which we can be viewed as "end of sequence" (EOS) token), we sample a token $\s$ from the distribution $\m(\p,\ss)$ (using a decoding algorithm, such as multinomial sampling, or greedy sampling of the single most likely next token)  and append it to $\ss$. The algorithm stops once a special token $\bot$ is emitted. Once the decoding algorithm is fixed, we can model $\m$ as taking a prompt $p$ in $\V^\star$ and outputting a string in $V^\star$. In a slight abuse of notation, we will henceforth use $\m(\p)$ to denote the response string of the \LLM~on the input prompt $\p$.\\

\subsubsection{Tokens and Types}
Given a sequence of tokens $\ss \in \V^{\star}$, a {\it typed sequence} is a $2$-tuple  $\ss_{\tau}=\langle (\sigma_1,\tau_1), \cdots, (\sigma_n, \tau_n) \rangle$,
where  $\tau_i \in \T $ is the type of the substring $\sigma_i$ of $\sigma$
(we assume $\ss = \sigma_1 \cdot \sigma_2 \cdots \sigma_n$). Each type is associated with a domain.
We also assume the existence of a \textit{type annotator}.
\begin{defn}[Type Annotator]\label{def:AT} A type annotator is a deterministic algorithm $\AT:\V^*\mapsto (\V^\star \times \T)^*$ that inputs a prompt $\p$  and outputs the corresponding typed sequence $\langle (\s_1,\t_1),\cdots,(\s_n,\t_n)\rangle$. 
\end{defn}
For example, consider the 
following prompt $\p$: ``Kaiser Soze is 50 years old and earns 500,000 per year. What is his ideal retirement plan?'' $\AT(\p)$ is  given as follows: ``(Kaiser Soze, [{\it Name}]) is (50, [{\it Age}]) years old and earns (500,000 , [{\it Salary}]) per year. What is his ideal retirement plan?'', where [{\it Name}], [{\it Age}], [{\it Salary}] are types of the tokens that precede it. For the ease of notation, here we only annotate tokens with sensitive types, i.e., all other non-annotated tokens have type $\perp$ (which
denotes non-sensitive token).
Note that type annotation is context dependent. For example, consider the following two prompts: 
$\rho_1$= ``My age is 53 years.'' and $\rho_2$= ``I stay at 53 Broadway Street.'' The same token $53$ has two different types in the two prompts: type \textit{Age} and type \textit{Street Number} in $\rho_1$ and $\rho_2$, respectively.

%% file: NDSS_revision/3_ProposedScheme.tex
\section{Prompt Sanitizer}
\label{sec:scheme}

Given an input prompt $\p$, a \textit{prompt sanitizer} (denoted by $\PS$) transforms the entire prompt to a sanitized one $\hat{\p}$ with the goal of protecting the sensitive tokens contained in $\p$. 
It is formally defined as follows:

\begin{defn}[Prompt Sanitizer] A prompt sanitizer $\PS=\langle \Set,\AT,
\E,\D\rangle$ is a tuple of the following algorithms:

\begin{itemize}\item \textit{Setup} ($\Set$).  The setup algorithm takes no input and outputs a secret key, as $\K\leftarrow \Set$. 
\item \textit{Type Annotator} ($\AT$). The type annotator inputs a prompt (token sequence) $\p\in\V^*$ and outputs the corresponding type-annotated token sequence as $\p_{\tau}\leftarrow \AT(\p)$ (as defined in Def. \ref{def:AT}).
\item \textit{Sanitization} $(\E)$. The sanitization algorithm  takes as input the secret key $\K$ and a 
type-annotated token sequence $\p_{\tau} \in (\V^\star \times\T)^*$. It outputs a token sequence $\hat{\p} \in \V^{|\p_{\tau}|}$, as $\hat{\p}\leftarrow \E(\K, \p_{\tau})$. 
\item \textit{Desanitization} $(\D)$. Desanitization takes a string (token sequence) $\hat{\r} \in \V^*$ and processes it with the goal of reversing the effect of the sanitization algorithm, using the
secret key $\K$. This is represented as $\r\leftarrow \D(\K, \hat{\r})$ with $\r \in \V^{|\hat{\r}|}$. 
\end{itemize} 
\label{def:PS}
\end{defn}
Given a prompt $\p$, the typical workflow of \PS~proceeds as follows:\\
$(1)$ type annotate the prompt to obtain $\p_{\tau}=\AT(\p)$, \\
$(2)$ sanitize the type annotated prompt using the secret key $\K$ as $\hat{\p}=\E(\K,\p_{\tau})$, \\
$(3)$ obtain the \LLM's response on the sanitized prompt as $\hat{\r}=\m(\hat{\p})$,\\
$(4)$ desanitize the response to obtain $\r=\D(\K,\hat{\r})$. \\

In the above workflow, the desanitization algorithm restores information about the original prompt $\p$ in its output, $\r$. In the special case where we run the desanitization algorithm directly on the sanitized prompt $\hat{\p}$ (which can be useful for instance if the \PS~is used to store a set of sensitive prompts on an untrusted platform for later use), we ideally expect $\r=\p$. 

We require that the sanitization and desanitization algorithms  are \textit{type preserving}, which means that if $\p =\langle \s_1, \cdots, \s_n\rangle$ and $\p_{\tau} = \langle (\s_1,\t_1),\cdots,(\s_n,\t_n)\rangle\leftarrow \AT(\p)$ and $\hat{\p}\leftarrow \E(\K, \p_{\tau})$ and $\langle (\s_1',\t_1'),\cdots,(\s_n',\t_n')\rangle\leftarrow \AT(\hat{\p})$ then it must be that $(\t_1,\cdots,\t_n)=(\t_1',\cdots,\t_n')$. 

\subsection{Privacy Guarantee}
\label{subsec:privacy-game}
The privacy game, denoted as $\gamePS{\PS, \L}$, is designed to capture an adversary's ability to distinguish between the sanitized outputs of two different prompts. In the game if the adversary picks two prompts that have a very different structure (e.g. different type or number of tokens), then the adversary can trivially distinguish between the corresponding sanitized prompts. To rule out pathological cases, we restrict the adversary to selecting two prompts with a ``similar structure'', formalized via a leakage function.
Different instantiations of the leakage function lead to different instantiations of the game.

The game is defined as follows:

\begin{algorithmic}[1]
\Statex \underline{$\Initialize$}: 
\State $\K \leftarrow \Set()$ 
\State $b  \overset{{\scriptscriptstyle\$}}{\leftarrow} \bits$ \textcolor{blue}{$\rhd$} Select a random bit 
\Statex \underline{$\Sanitize(\p_0, \p_1)$}: \textcolor{blue}{$\rhd$} Adversary selects two prompts 
\State $L_0 \gets \L(\p_0)$ ; $L_1 \gets \L(\p_1)$  

\Statex \textcolor{blue}{$\rhd$}  $\L$ is the leakage  function associated  with $\PS$
\State \textbf{if} $L_0 \neq L_1$ 
\textbf{then return} $\bot$
\Statex \textcolor{blue}{$\rhd$} Only prompt pairs with the same leakage   are valid 
\State $\hat{\p}_0 \gets \E(\K, \AT(\p_0))$ ; $\hat{\p}_1 \gets \E(\K, \AT(\p_1))$  
\State \textbf{return} $\hat{\p}_b$ 
\Statex \textcolor{blue}{$\rhd$} Return one of the sanitized prompts chosen at random
\Statex \underline{$\Finalize(b')$}:

\State \textbf{return} $[b' = b]$ 
\hfill \textcolor{blue}{$\rhd$} $b'$ is the adversary's guess for $b$
\end{algorithmic}
We denote an adversary by $\advppt$. We model the information leakage from the sanitized prompts through a leakage function,  $\L$. In particular, the leakage function $\L$ of a prompt sanitizer takes as input a prompt $\p$ and captures all the information about sensitive tokens that is leaked by $\hat{\p}=\E(\p,\K)$, given a key $\K$. In the above game, an adversary $\advppt$ aims to distinguish between the sanitized prompts of $\p_0$ and $\p_1$ based solely on the sanitized output and the leakage allowed by $\L$. The adversary is said to win the game if $b'=b$ and their advantage is formally defined as:
\begin{align*}
\vspace{-0.2cm}
\ppAdv{\PS, \L}{\advppt} = 2\Pr[\gamePS{\PS, \L}(\advppt) = 1] - 1. \vspace{-0.5cm}
\end{align*}
Intuitively, the game implies that even after observing a sanitized prompt, an adversary should not be able to reliably differentiate between two prompts with the \textit{same} leakage. The definition of the leakage function, $\L$, is crucial and depends on the underlying sanitizer $\E$. For example, if $\E$ sanitizes a token by redaction, for prompt $\p = $``My age is 26'', the leakage function outputs all the non-sensitive tokens, i.e., $\L(\p)$ = ``My age is [ ]''—this represents the minimal possible leakage as redaction is the strongest sanitization mechanism. Alternatively, if $\E$ encrypts sensitive tokens, $\L$ might reveal the length of these tokens to $\advppt$. Note that leakage function is a standard notion in cryptography~\cite{dziembowski2008leakage}. For instance, the leakage function for order-preserving encryption~\cite{frequencyHiding2} is essentially the numerical ordering of the input dataset. 

The restriction in the above game, requiring the pair of prompts to have the same leakage, aligns with standard notions in game-based cryptographic security definitions. For instance, this is similar to the definition of security in order-preserving encryption (IND-FA-OCPA~\cite{frequencyHiding2}), where the adversary is restricted to selecting pairs of data sequences that maintain the same order. 

 \subsection{Utility Guarantee}
   Let $\Q: \V^* \times \V^* \mapsto R_{\Q}$ be a \textit{quality oracle} that evaluates the quality of a candidate response $\r$ for a prompt $\rho$. Specifically, $\Q(\p,\r)$ is a measure of the response's goodness. Such a quality oracle has been used in prior work on \LLM s~\cite{zhang2024watermarks}.
\begin{defn}
    A prompt sanitizer  $\PS$~satisfies $(\alpha,\beta)$ utility for a given prompt $\p \in \V^*$,
    \begin{gather}
    \alpha = \mathbb{E}_{f}\Big[Q\big(\p,f(\p)\big)\Big]\\
    \beta = \mathbb{E}_{f,\PS}\Big[Q\Big(\p,\D\big(\K,f(\hat{\p})\big)\Big)\Big], \hat{\p} = \E\big(\K,\AT(\p)\big)
    \end{gather}
    where the randomness is defined over both the LLM, \m,~ and \PS.
\end{defn}
The utility of the prompt sanitizer \PS~is evaluated by comparing the quality of the original response $f(\p)$ with the one obtained through the \PS~pipeline. The above definition has two key characteristics. First, the utility is defined w.r.t to a specific prompt, as response quality can vary significantly across different prompts. For example, consider the following prompt: $\p$ = \textit{``My age is 46. What is the average age of the population of New York?''}
Here, a high-quality \LLM's response should be invariant to the sensitive token ([\textit{Age}]) in the prompt. This means that even after sanitization, we should be able to retrieve a correct and relevant response. On the other hand, for a conversational prompt used in a \LLM-based chatbot to seek medical advice, the quality of the responses could vary significantly based on the specifics of the sanitization and desanitization algorithms of \PS. Note that the quality oracle $Q$ can take various forms based on the type of the prompt. For instance, it might be a human evaluator who assigns a quality score, or it could be a predefined analytical expression in case the prompt has some special structures. 
Second, utility is defined as an expectation, since in general both the \LLM~\m~and the prompt sanitizer \PS, are probabilistic. 
Note that when the distribution of $f(\p)$ matches the distribution of $\D(\K,\m(\hat{\p}))$, this represents the strictest form of utility. If $R_Q$ is a metric space with a distance metric $d_Q$, we can quantitatively measure the mean degradation in the quality of the response as $d_Q(\alpha, \beta)$.

%% file: NDSS_revision/4_Preempt.tex
\section{\texorpdfstring{\name}: System Description}\label{sec:name}
This section introduces \name: a system that instantiates a sanitizer for prompts.
First, we describe the primitives (FPE and mDP) in subsection~\ref{sec:primitives}.
Next, two sections describe our threat model and design goals. Our system is described
in subsection~\ref{sec:name:modules}. We conclude the section with privacy and utility analysis.
Our last subsection discusses other ``strawman'' solutions and why they don't address our threat
model and goals.

\subsection{Building Blocks}
\label{sec:primitives}
We start by describing the building blocks which will be used to sanitize the sensitive tokens. 

\subsubsection{Format-Preserving Encryption (\FPE)}\label{sec:background:FPE}
Under a format preserving encryption (\FPE) scheme, the plaintext and the ciphertext have the same format, that is, FPE ensures that the encrypted output is structurally similar to the original input, including properties such as length, character set, or format. This property allows applications to process ciphertexts and plaintexts in the same way. This backward compatibility makes FPE a popular tool for secure data analytics in practice. For instance, the ciphertext of a 16-digit credit card number encrypted under a \FPE~scheme would also be a 16-digit number\footnote{One can add additional constraints, such as ensuring the last digit is the Luhn checksum or the first four digit corresponds to a bank.}. As concrete examples, a plaintext Social Security number such as 055-46-6168 might be transformed into the ciphertext 569-83-4469, while an IP address like 76.217.83.75 could be encrypted as 97.381.64.35. Intuitively, FPE shuffles or re-encodes values within the space of all valid values of the same type. Without the decryption key, the ciphertext looks indistinguishable from any legitimate value of that format, effectively hiding the original data while remaining compatible with systems that expect specific formats.


\begin{defn}[Format Preserving Encryption (\FPE)] A format preserving encryption scheme is a tuple $\mathcal{E}=\langle \GenF,\EF,\DF \rangle$ of polynomial time algorithms:
\begin{itemize}
\item \textit{Key Generation} ($\GenF$).  The key generation algorithm is probabilistic polynomial time algorithm that takes as input a security parameter $\kappa$ and outputs
a secret key $\K$ as $\K\leftarrow \GenF(1^\kappa)$. 
\item \textit{Encryption} ($\EF$)\footnote{\FPEs~also take a \textit{tweak space} as an input which we omit here for the ease of exposition}. The encryption algorithm  is deterministic polynomial time algorithm that takes as input a secret key $\K$, a
plaintext $x \in \M$, and a format $\textsf{N} \in \Nf$ and outputs a ciphertext $y \in \M$ as $y\leftarrow \EF(\K,\N,x)$. 
\item \textit{Decryption} ($\DF$). The decryption algorithm  is deterministic polynomial time algorithm that recovers the plaintext as $x\leftarrow \DF(\K,\N,y)$.
\end{itemize}
\end{defn}

Typically, the format of a plaintext is described as a finite set $\N$ over which the encryption function induces a permutation. For example, with SSNs this is the set of all nine-decimal-digit numbers.

\subsubsection{Metric Local Differential Privacy (\mLDP)}\label{sec:background:metricDP}

Differential privacy (\DP) is a quantifiable measure of the stability of the output of a randomized mechanism to changes to its input. As a direct consequence of our threat model (Sec. \ref{sec:name:threat}), we work with the local model of DP (\LDP) where each data point is individually randomized. Metric local differential privacy (\mLDP)~\cite{chatzikokolakis2013broadening, Alvim2018, OPe, MetricDP} is a generalization of \LDP~which allows heterogenous guarantees for a pair of inputs based on a distance metric $d(\cdot)$ defined over the input space. 
\begin{defn}[Metric Local Differential Privacy (\mLDP)~\cite{chatzikokolakis2013broadening}]
 A randomized algorithm 
 $\mathcal{M} : \mathcal{X} \rightarrow \mathcal{Y}$ is $\epsilon$-\mLDP~for a given metric $d: \mathcal{X}\times \mathcal{X} \mapsto \mathbb{Z}_{\geq 0}$ if for any pair of private values $x, x' \in \mathcal{X}$ and any subset of output, $\mathcal{O} \subseteq \mathcal{Y}$ \begin{gather}
 \mathrm{Pr}\big[\mathcal{M}(x) \in \mathcal{O}\big] \leq e^{\epsilon d(x,x')} \cdot \mathrm{Pr}\big[\mathcal{M}(x') \in \mathcal{O}  \big]\end{gather}
\end{defn}
\noindent
\mLDP~uses the distance between a pair of values to customize heterogeneous (different levels of)
privacy guarantees for different pairs of private values. In particular, the privacy guarantee degrades linearly with the distance between a pair of data points; that is, only data points that are ``close'' to each other should be indistinguishable. Still, \mLDP~captures the privacy semantics of many real-world scenarios and is well suited to settings where releasing approximate information is acceptable. For example, it is often sufficient to reveal a coarse location, such as a city block, rather than exact GPS coordinates. Similarly, sharing an income range (e.g., \$60K–\$80K) can preserve utility without exposing precise figures.
Alg. \ref{alg:Mdp} in App.~\ref{app:emldp-algo}, outlines a mechanism for achieving $\epsilon$-\mLDP~for the $\ell_1$ distance using a variant of the exponential mechanism \cite{Dwork}. 

\begin{theorem}
Mechanism $\Mdp$~satisfies $\epsilon$-\mLDP~for the $\ell_1$ distance. \label{thm:mLDP}
\end{theorem} 
The proof of the theorem is standard and appears in the appendix.
However, we next justify why $\Mdp$ is an appropriate notion for
our context. An input is more likely to be mapped to one which is close to it, which we formalize this using the following two properties. 
\\
\textbf{Property 1.}
\begin{gather*}
\mathrm{Pr}\big[\Mdp(x,\epsilon,[k]) = x\big]> \mathrm{Pr}\big[\Mdp(x,\epsilon,[k])= y\big], \forall y\in [k]\label{eq:mdp:order:1}
\end{gather*}  
\textbf{Property 2.}
\begin{gather*} |y_1 -x| < |y_2-x| \iff\\
\mathrm{Pr}\big[\Mdp(x,\epsilon,[k]) = y_1\big]> \mathrm{Pr}\big[\Mdp(x,\epsilon,[k])= y_2\big], \\\forall y_1,y_2 \in [k] 
\end{gather*}

\subsection{Threat Model}\label{sec:name:threat}
\name~runs as an application on a user's (trusted) local device. Additionally, \name~can support multiple users: envision it as an application maintained at the level of an organization and available to all of its employees. The user inputs a string $(\V^*)$ to \name~and obtains a transformed string. Every such interaction constitutes a \textit{separate} session. In particular, consider the following chain of events. An user $\U$ submits a prompt $\p$ to \name~and obtains a sanitized version of it $\hat{\p}$. Next, they obtain a response $\hat{\r}$ from an \LLM~on $\hat{\p}$ and again uses \name~to desanitize it into $\r$. This interaction constitutes two separates \name~sessions: one for the $\p \to \hat{\p}$ transformation and the other for the $\hat{\r}\to \r$ transformation. The \LLM~is an untrusted third-party application which represents the adversary (Fig. \ref{fig:overview}). 

In \name, we focus on tokens where
the sensitive information can be derived \textit{solely} from the individual token, with no extra context. Examples of such token types include SSN, credit card number, license number, age, money, A/C number, zipcode.  Privacy issues stemming from the linguistic context of the prompts (e.g. a prompt indicating the users’ mental health details) are beyond \name's scope (see App. \ref{sec:Open} for more discussion).

\subsection{Design Goals}
\label{sec:name:design}
\name~has the following design goals.

\squishlist\item \textbf{Formal Guarantees.} \name~should be able to provide a formal privacy guarantee on the sanitized prompts. 

\item \textbf{High Utility.} The responses based on sanitized prompts should be ``close'' to the responses based on the original prompt. 

\item \textbf{Stateless.} Finally, the sanitization and desanitization process should be stateless, i.e., \name~should not retain information (state) from any prior session. This design choice offers dual advantages. Firstly, storing sensitive information derived from users' prompts/responses post-session termination would violate privacy and contravene legal frameworks, such as the EU's GDPR~\cite{GDPR} and California's CCPA~\cite{ccpa}. Additionally, these regulations grant individuals the Right to Deletion, allowing data owners to retract authorization previously granted for the use of their personal data. A stateful solution hinders the Right to Deletion and desanitization, while a stateless one offers flexibility and storage efficiency. For example, consider these two user action sequences:
\vspace{-0.2cm}\begin{align*}
    A_1= \langle  \mbox{ Sanitize } \p_1; \mbox{ Desanitize } \hat{\r}_1; \mbox{ Sanitize } \p_2 ; \\ \mbox{ Desanitize } \hat{\r}_2; \mbox{ Sanitize } \p_3; \mbox{ Desanitize } \hat{\r}_3 \rangle \\ A_2= \langle \mbox{ Sanitize }\p_1; \mbox{ Sanitize } \p_2; \mbox{ Desanitize } \hat{\r}_2,\\ \mbox{  Sanitize } \p_3,  \mbox{ Desanitize } \hat{\r}_1, \mbox{ Desanitize } \hat{\r}_3\rangle. \vspace{-0.2cm}\end{align*}

 Without perpetual retention of state information, a stateful solution restricts a user to a specific action sequence of sanitizing and desanitizing \textit{in order} (such as, $A_1$). Moreover, multiple desanitization of the \textit{same} string cannot be supported without perpetual storage of the state information.  The issue is exacerbated with multiple users as a stateful solution entails storing separate state information for each user.  In contrast, a stateless solution provides the flexibility of supporting arbitrary sequences of user actions (such as, $A_2$). 
\squishend

Note that while \name~is stateless, conversation with the LLM can be stateful -- the LLM is free to maintain a history of all (sanitized) prompts to better respond to user queries. We illustrate this experimentally in Sec. \ref{sec:experiments}.

\subsection{System Modules} 
\label{sec:name:modules}

\name~supports three types of sessions, namely, User Registration, Sanitization and Desanitization, which is taken care of by the following modules.
\\\noindent
\textbf{Configuration Manager.} The configuration manager module of \name~generates a secret key $\K_{\U} \rightarrow \Gen(1^\kappa)$ for a given security parameter for an user $\U$ at the time of the registration for a \FPE~scheme $\mathcal{E}$. Subsequently, for any session involving user $\U$, this module initializes all instances of the \FPE~scheme with  the key $\K_{\U}$. Additionally, during registration, user $\U$ specifies the privacy parameter $\epsilon$ for $\Mdp$, which is treated as the privacy budget for each individual sanitization session. The module also initializes the data domain (equivalently, format in the case of \FPE) for each sensitive token type.  The domains can either be predefined or computed based on some user-provided information. Lastly, various parameters (e.g. format and privacy parameter) can be dependent on the type $\tau$. 

\noindent\textbf{Sanitizer.} Recall that \name~only sanitizes sensitive tokens that are alphanumeric or numeric (see Sec. \ref{sec:name:threat}). To this end, \name~assumes that such sensitive tokens fall into two distinct categories:
\squishlist 
    \item \textit{Category I $(\tau_{\texttt{I}})$.} These tokens are characterized by the fact that the \LLM's response depends \textit{solely} on their format. Examples of tokens in this category include names, Social Security Numbers (SSN), credit card numbers, Taxpayer Identification Numbers (TIN), passport numbers, bank account numbers, driver's license numbers, phone numbers, license numbers, and IP addresses. We provide empirical evidence of this assumption in Sec.~\ref{sec:experiments}. 
    \item \textit{Category II $(\tau_{\texttt{II}})$.} This category encompasses tokens where the \LLM's response hinges on the specific \textit{numerical value} itself, such as age, medical records\footnote{Note that in some cases, for instance, language translation, all sensitive tokens are of $\t_{\texttt{I}}$ since the \LLM's response should not depend on the specific values.}, etc. That is, the \LLM~performs specific computations based on the \textit{values} of these tokens. 
\squishend

\begin{algorithm}[thbt]
 \caption{\name: Sanitization }

\begin{algorithmic}[1]
\Statex \textbf{Input:} $\p$ - Input prompt;  $\K_{\U}$ - Sanitization key; $\epsilon$ - Total budget 
\Statex \textbf{Output:} $\hat{\p}$ - Sanitized prompt;
\State $\p'=\langle \rangle$
\State $\p_{\tau}\leftarrow \AT(\p)$ \textcolor{blue}{$\rhd$} $\AT$ is instantiated with a named-entity recognizer
\State $(\psi,t) \leftarrow \Mpre(\p,\p_{\tau})$ \textcolor{blue}{$\rhd$} $\Psi$ is a helper string encoding some extra information about the type of tokens
\Statex \textcolor{blue}{$\rhd$}  $t$ is the number of tokens in $\p_{\tau}$ with type $\tau_{II}$
\State \textbf{for} $(\s,\tau) \in \p_{\tau}$
\State \hspace{0.3cm} \textbf{if} $(\tau \neq \NS)$
\State \hspace{0.6cm} \textbf{if} $(\tau == \tau_{\texttt{I}})$
 \State \hspace{0.9cm} $\hat{\s}=\EF(\K_{\U},\N_{\tau},\sigma)$ \textcolor{blue}{$\rhd$} $\N_{\tau}$ is the format of $\sigma$
\State \hspace{0.6cm} \textbf{else}
\State \hspace{0.9cm} $\hat{\s}=\Mdp(\sigma,\frac{\epsilon}{t}, k_\tau)$  \textcolor{blue}{$\rhd$} $[k_{\tau}]$ is the domain of $\sigma$
\State \hspace{0.6cm} \textbf{end if}
\State \hspace{0.3cm} \textbf{else}
\State \hspace{0.6cm} $\hat{\s}=\s$
\State \hspace{0.3cm} \textbf{end if}
\State \hspace{0.3cm} $\p'.append(\hat{\s})$
\State \textbf{end for}
\State $\hat{\p}\leftarrow \Mpost(\p',\Psi)$  \textcolor{blue}{$\rhd$} Performs some post-processing on the sanitized tokens
\State \textbf{return} $\hat{\p}$

\end{algorithmic}\label{alg:PS}
\end{algorithm}
A prompt is sanitized as follows. 
We first perform type annotation of the different tokens via $\AT$. In addition to annotating the type of a token, $\AT$ also indicates its category. For \name, we instantiate $\AT$ with a named-entity recognizer (NER).  Next, \name~uses a pre-processor $\Mpre$ that takes  $(\p,\p_{\tau})$ as input and computes two things; $1)$ it determines the number of tokens belonging to the second category, denoted as $t$, $2)$ it computes a \textbf{helper string $\Psi$} to encode additional information about token types and provide flexibility during sanitization. Specifically, $\Psi$ captures functional dependencies between the tokens. 
We present examples involving the helper string in Sec.~\ref{sec:param} and App.~\ref{app:helper_string}.

In \name, each sensitive token is sanitized individually. In particular,  all tokens of the first category are sanitized using \FPE~with the user specific secret key $\K_\U$. On the other hand, all tokens of the second category are sanitized to satisfy $\frac{\epsilon}{t}$-\mLDP~using $\Mdp$, where $\epsilon$ is the privacy parameter for the standard DP guarantee and $t$ is the maximum distance between protected values. No operation is performed on tokens with non-sensitive types ($\t=\perp$).  Next, all sanitized tokens are concatenated and passed to a post-processor $\Mpost$. The $\Mpost$~ enforces the functional dependencies encoded in $\Psi$. Furthermore, in \name~only the determinant is perturbed, and the sanitized versions of the dependent tokens are derived from this noisy encoding. If $\Psi$ is empty (i.e. not functional dependencies are provided), then each token is handled independently and no functional dependencies are enforced, which can adversely affect utility. 
The full sanitization mechanism is outlined in Algorithm  \ref{alg:PS}. Steps 3-17 in Alg.  \ref{alg:PS} instantiate the sanitization algorithm $\E$ of the prompt sanitizer (Def. \ref{def:PS}).

\begin{algorithm}[thbt]
 \caption{\name: Desanitization}\label{alg:RdS}

\begin{algorithmic}[1]
\Statex \textbf{Input:} $\hat{\r}$ - Input sanitized response; $\K_{\U}$ - Sanitization key; 
\Statex \textbf{Output:} $\r$ - Desanitized response;
\State $\r=\langle \rangle$
\State $\hat{\r}_{\tau}\leftarrow \AT(\hat{\r})$ \textcolor{blue}{$\rhd$} $\AT$ is instantiated with a named-entity recognizer
\State \textbf{for} $(\s,\tau) \in \hat{\r}_{\tau}$
\State \hspace{0.5cm} \textbf{if} $(\tau == \tau_{I})$
 \State \hspace{0.8cm} $\s=\DF(\K_{\U},\N_{\tau},\sigma)$ \textcolor{blue}{$\rhd$} $\N_{\tau}$ is the format of $\sigma$
\State \hspace{0.5cm} \textbf{else}
 \State \hspace{1cm} $\s=\hat{\s}$
\State \hspace{0.5cm} \textbf{end if}
\State \hspace{0.5cm} $\r.append(\sigma)$
\State \textbf{end for}
\State \textbf{return} $\r$

\end{algorithmic}
\end{algorithm}

\noindent\textbf{Desanitizer.} Desanitization (Alg. \ref{alg:RdS}) begins with the same type annotator. All sensitive tokens of category I can be desanitized using the decryption algorithm of the \FPE~scheme. However, tokens sanitized with $\Mdp$ cannot be desanitized without retaining additional state information and are hence, left untouched by default. Steps 3-11 in Algorithm  \ref{alg:RdS} correspond to the desanitization algorithm $\D$ of the prompt sanitizer (Def. \ref{def:PS}).

One drawback of this approach is that tokens from the first category that did not appear in the original prompt (and consequently were never sanitized) might also undergo desanitization.  Users can mitigate this by providing the original prompt $\p$ as auxiliary information. In this scenario, \name~will exclusively desanitize tokens that appeared in the prompt. Note that the only thing required to desanitize is the secret key $\K_\U$: \name~does not store any sensitive information post the termination of a session thereby making our solution stateless. 
\subsection{Setting up Parameters\label{sec:param}}
In this section, we provide guidelines on how to choose the parameters for \name.

\textit{Setting up $\epsilon.$} If a user desires a privacy parameter of $\epsilon$ for the standard DP guarantee and wishes to protect values that differ by at most distance $l$, then use $\epsilon'$-\mLDP~with $\epsilon' = \frac{\epsilon}{l}$. Any $\epsilon$-\mLDP~protocol is $\epsilon\cdot l_{max}$-DP where the distance between any pair of inputs is at most $l_{max}$.

\textit{Using the helper string $\Psi.$} 
The helper string $\Psi$ encodes auxiliary knowledge that captures functional dependencies among sensitive tokens. These dependencies typically fall into two categories:  
(1) \textit{Common knowledge}, such as mathematical or definitional identities (e.g., $\textit{Annual Salary} = 12 \times \textit{Monthly Salary}$), and  
(2) \textit{Domain-specific knowledge}, such as medical or financial formulas (e.g., $\textit{BMI} = \frac{\textit{Weight}}{\textit{Height}^2}$).

For common knowledge, the dependencies can be stored in a knowledge base and appended directly to the prompt to aid reasoning or consistency. Since these are standardized and widely accepted relationships, they can be programmatically injected with little ambiguity.

Handling domain-specific dependencies is more complex. One possible approach is to use a local LLM to infer dependencies between sensitive attributes identified by NER, producing structured representations of causal or computational links. These dependencies can then be modeled as a directed acyclic graph (DAG), where each node corresponds to a sensitive attribute and edges represent dependencies (e.g., computation or inference). The root nodes represent the base sensitive values and are directly noised using \mLDP. These values are then propagated along the DAG via the encoded dependencies, ensuring that related fields (such as income and tax, or weight and BMI) are sanitized consistently.
\subsection{Privacy and Utility Analysis}
\label{sec:privacy-and-utility-analysis}

\noindent\textbf{Privacy Analysis.} The formal privacy guarantee of \name~is given as follows:
\begin{theorem} Let  $S$ be the set of all token pairs of type $\tau_{\texttt{II}}$ that are different in the prompt pairs $(\p_0,\p_1)$ in the privacy game $\gamePS{\PS, \L}$. Then, for 
\name~we have:
\begin{gather}
\ppAdv{\text{Pr}\epsilon\epsilon \text{mpt}, \L}{\advppt}\leq e^{l \epsilon} + negl({\kappa}) \label{eq:privacy}
\end{gather}
 where $l=\max_{(\sigma_0,\sigma_1)\in S}\{|\sigma_0-\sigma_1|\}$ and  $\kappa$ is the security parameter of the underlying \FPE~scheme. 
 \label{thm:privacy}
\end{theorem}
\noindent\textit{Proof Sketch.}  First, we compute the adversary's advantage in \name~when the two prompts $(\p_0,\p_1)$ differ by only a single token, denoted as $\ppAdv{\text{Pr}\epsilon\epsilon \text{mpt}, \L=1}{\advppt}$. Next, using the classic hybrid argument~\cite{Oded}, we establish an upper bound on the adversary's advantage in the general case, expressed in terms of its advantage when the prompts differ by just a single token. Finally, Eq. \ref{eq:privacy} can be derived by substituting $\ppAdv{\text{Pr}\epsilon\epsilon \text{mpt}, \L=1}{\advppt}$ into this result. The full proof is presented in Appendix~\ref{app:proof}. 

\noindent \textbf{Practical Privacy Considerations.}\\{\noindent\textit{Error due to NER.} \name’s privacy guarantee is cryptographic and \textit{orthogonal} to NER. Specifically, \name~uses NER as a black-box and the above theorem aligns with the standard $\mathcal{F}_\text{NER}$-hybrid model of security where $\mathcal{F}_{\text{NER}}$ is an ideal functionality for NER~\cite{oded_goldreich}. Importantly, the performance of NER should \textit{not} be conflated with the efficacy of the sanitization scheme. While the practical performance depends on properly identifying sensitive tokens, this is inherent to our task. As NER continues to improve, so will Pr$\epsilon \epsilon$mpt's  practical performance \textit{without} any modifications to the design. Additionally, domain-specific pattern matching with regular expressions can achieve high performance for structured data.} 

Nevertheless, we provide the following result that formalizes how to analyze privacy in the presence of NER errors.  The main idea is to model errors within the leakage function $\L$ of our privacy game $\gamePS{\PS, \L}$. If the false negative rate of the NER is $\lambda$, we define a leakage function $\mathcal{L}_\text{NER}$ that additionally leaks up to $\lambda\%$ of the sensitive tokens. To construct the two prompts $\p_0$ and $\p_1$ for the corresponding privacy game $\gamePS{\PS, \mathcal{L}_{\text{NER}}}$, we proceed as follows: we start with the original prompt pair as in the unmodified game $\gamePS{\PS, \mathcal{L}}$, and then modify $\p_1$ by replacing $\lambda\%$ of its sensitive tokens with the corresponding tokens from $\p_0$. In other words, this construction models the scenario where the adversary gains access to a fraction of sensitive tokens that were missed by the NER. \name~then protects the rest of the sensitive tokens as:
\begin{theorem} Let  $S$ be the set of all token pairs of type $\tau_{\texttt{II}}$ that are different in the prompt pairs $(\p_0,\p_1)$ in the privacy game $\gamePS{\PS, \L_{\text{NER}}}$. Then, for 
\name~we have:
\begin{gather}
\ppAdv{\text{Pr}\epsilon\epsilon \text{mpt}, \L_{\text{NER}}}{\advppt}\leq e^{l \epsilon} + negl({\kappa}) 
\end{gather}
 where $l=\max_{(\sigma_0,\sigma_1)\in S}\{|\sigma_0-\sigma_1|\}$ and  $\kappa$ is the security parameter of the underlying \FPE~scheme. 
 \label{thm:privacy:NER}
\end{theorem} The proof of the above theorem is in App. \ref{app:thm:NER}.

\textit{Correlated Tokens.}  To mitigate correlation attacks, \name~adopts a conservative approach by dividing the privacy budget equally among \textit{all} tokens of type $\t_{\texttt{II}}$. This ensures that, via composition, the total privacy loss remains bounded by $\epsilon$. However, leveraging the helper string $\Psi$ can yield a better privacy-utility tradeoff. For example, in the prompt, “My age is $X$, I was born in $Y$. I am X years old.”, $[Age: X]$, $[Year: Y]$ and $[Age: X]$ are the sensitive tokens.
By default, \name~distributes the privacy budget equally ($\epsilon$/3) among all type $\t_{\texttt{II}}$ tokens, however, the helper string $\Psi$ can indicate that $X$ and $Y$ represent the same ground-truth and that $X$ is repeated. Using $\Psi$, \name~applies $\epsilon$-\mLDP~to the first occurrence of $X$, yielding $\hat{X}=25$ (suppose). \name~then derives the corresponding $\hat{Y} = 2000$ by post-processing and reuses $\hat{X}$ for the second occurrence of age. This incurs no additional privacy loss due to the post-processing immunity of \mLDP~\cite{10.1561/0400000042}. Thus, the resulting sanitized prompt is given by: “My age is 25, I was born in 2000. I am 25 years old.”. We present additional illustrative examples in App.~\ref{app:helper_string}. \\

\noindent\textbf{Utility Analysis.}  We analyze \name's utility below.
\\\noindent\textit{Prompts with Perfect Utility.} Recall  that, for assessing utility, we compare the responses of the \LLM~to the original prompt $\p$ and the sanitized prompt $\hat{\p}$ produced by \name. For many practically useful prompts, the response of the \LLM~remains the \textit{same} for both $(\p,\hat{\p})$ except for the substitution of the sensitive tokens $\s \in \p$  with  their sanitized counterparts  $\hat{\s}$. In other words, the sanitized response $\hat{\r}$ generated from $\hat{\p}$ preserves perfect utility (after desanitization).  We refer to such prompts as \textit{invariant prompts}, where the \LLM's response should be invariant to the specific values (or small variations) of the sensitive tokens. This property holds in particular for prompts containing only type $\t_{\texttt{I}}$ tokens. Translation is exemplar: all sensitive tokens will be classified as type $\t_{\texttt{I}}$ and sanitized using \FPE~since the \LLM's translation should not depend on their specific values. 
As a result, sanitized tokens can be perfectly desanitized from the translated text. The quality score (the output of $Q$) can be evaluated using metrics such as BLEU score~\cite{BLEU}.

We now turn to the case of invariant prompts that include sensitive tokens of type $\t_{\texttt{II}}$. One such example is a factual information retrieval task for RAG. Consider the following prompt in the context of financial documents: ``Please return all bank accounts with balance greater than \$2000.'' Here the two sensitive tokens [\textit{Bank A/C}] and [\textit{Bank Balance}] are sanitized via \FPE~and \mLDP, respectively. \mLDP, by construction, noisily maps an input to a value that is close to it (as per Properties 1 and 2 in Sec. \ref{sec:background:metricDP}).  As a consequence, the bank balance is perturbed only slightly, allowing correct numeric comparisons with high probability. This is precisely the rationale for our choice of \mLDP: sanitized tokens preserve ordinal relationships and remain close to their original values, enabling useful computations while still providing strong privacy guarantees. The quality score here is the accuracy of the answers (count of the correct bank A/Cs returned). The above discussion is validated by our  experimental results in Sec. \ref{sec:experiments}. Formally, we have:

\begin{theorem} For invariant prompts, \name~satisfies $(\alpha, \alpha)$- utility where $\alpha =\mathbb{E}_{f}\Big[Q\big(\p,f(\p)\big)\Big]= \mathbb{E}_{f,Pr\epsilon\epsilon mpt}\Big[Q\Big(\p,\D_{Pr\epsilon\epsilon mpt}\big(\K,f(\hat{\p})\big)\Big)\Big]$.
\end{theorem}
\noindent\textit{Other Prompts}. Given the complex and open-ended nature of prompts and responses, it is challenging to assign a utility score for any general prompt. Nevertheless, we provide some guidelines for when \name~is likely to perform well. Note that \name~introduces only small perturbations in the sanitized prompt $\p$. Hence, intuitively, \name~should perform well where small changes in the original prompt result in only \textit{limited} changes to the generated response. There can be two natural ways to capture these changes in the response. 
First, consider cases where the prompt satisfies Lipschitz continuity~\cite{NoceWrig06}, as given by $ d_{\V}(\m(\p),\m(\hat{\p})) \leq 
K d_{\V}(\p,\hat{\p})$ for some $K \in \mathbb{R}_{>0}$ and distance metric $d_{\V}: \V^*\times \V^*\mapsto\mathbb{R}_{>0} $. Distances defined over a document embedding space could be apt for $d_{\V}$. For example, when using an \LLM~as a financial advisor with a prompt ``My monthly salary is \$12,000. Suggest a monthly savings plan.'', the response should ideally remain consistent (and hence, very close in the embedding space) even if the salary value is slightly altered to \$11,500 (via $\Mdp$).  
A second way of bounding changes in the response is when the operations of the \LLM~on the sensitive tokens can be expressed as a symbolic computation. For example; ``My height is 158 cm and weight is 94lb. Compute my BMI.'' The BMI is computed via a fixed formula (i.e., a symbolic computation). These type of prompts ensure that the responses on $\hat{\p}$ can deviate from the original response in only a well-structured and predictable manner. Additionally, if this symbolic mapping is known, \name~could leverage this information during desanitization to improve utility further. 
\\
\textit{Usability.} 
A key advantage of \name~is its ease of use: after type annotation, \name~employs predefined sanitizers to protect sensitive tokens without any manual configuration of custom rules or execution of ad hoc sanitization strategies. In the current prototype, $\Psi$ is treated as an optional user input. Importantly, while $\Psi$ can be used to improve performance—for example, through better privacy budget allocation—omitting it does not affect the privacy guarantees of \name. As discussed in Sec.~\ref{sec:param}, $\Psi$ can also be automatically generated from the input prompt to capture functional dependencies between sensitive tokens, further enhancing usability.

\subsection{Comparison with Strawman Solutions} 

\noindent\textbf{Strawman Solution I: Redaction.} One intuitive solution is to redact all sensitive tokens from the prompt. While this approach ensures perfect privacy, it severely impacts utility dependent on those sensitive tokens and can sometimes lead to a complete loss of functionality. We present an example in Sec.~\ref{appendix:strawman}

\noindent\textbf{Strawman Solution II: Substitution.} An intuitive solution is to replace sensitive tokens with others of the same type using a lookup table. However, this method is not stateless, as desanitization requires access to the lookup tables, leading to scalability and security issues (see Sec.~\ref{sec:name:design}). The table size grows linearly with the number of sensitive tokens, and a separate table is needed for each user to prevent information leakage. In contrast, \name~only requires a fixed-size key per user, regardless of prompt number or length.


\noindent\textbf{Strawman Solution III: Suppression.}
Consider the following  sanitization strategy for the tokens of the second category $(\t_\texttt{II})$: a numerical token is sanitized by simply setting its $k$ lowest order digits to 0, the intuition being that the \LLM's response is most likely to depend on the higher-order digits, thereby preserving utility while only leaking information about the numerical value at a coarser granularity.  However, it is difficult to formally quantify its privacy guarantees. Prior work shows that such ad-hoc approaches are often vulnerable to attacks~\cite{Dwork2017ExposedAS,279958,10.1145/1743546.1743558}.  In contrast, the \mLDP-based approach used in \name~offers a principled way of balancing this privacy/utility trade-off.

\noindent\textbf{Additional Baseline: LLMs assisted obfuscation and deobfuscation.} One could also attempt to use a LLM to obfuscate and deobfuscate sensitive information based on rules in the system prompt, and maintaining a state to recover information, such as~\cite{siyan2024papillon}. However, the probabilistic nature of the LLM and lack of specifications preclude any rigorous privacy analysis.

%% file: NDSS_revision/4_Evaluation_Short.tex
\section{Experiments} \label{sec:experiments}
We evaluate the  following questions:

\begin{tcolorbox} \vspace{-0.2cm}
\textbf{Q1.} How does \name~impact utility of realistic tasks compared to unsanitized performance?

\textbf{Q2.} How does \name~compare against prior LLM based sanitization approaches?

\textbf{Q3.} What is the impact of different technical design choices on \name's utility?  \vspace{-0.2cm}
\end{tcolorbox}

\subsection{Utility Loss from \texorpdfstring{\name}~ Sanitization}
\label{sec:setup}
We tackle \textbf{Q1} by applying \name~to four tasks: translation, retrieval-augmented generation (RAG), multi-turn financial question answering (Q/A), and long-context reading comprehension Q/A. These tasks represent a broad spectrum of real-world LLM applications where input prompts are likely to contain sensitive information.

\noindent\textbf{Models.} We use GPT-4o \cite{openai2023gpt4}, Gemini-1.5 \cite{geminiteam2024gemini15unlockingmultimodal}, and OPUS-MT for translations, RAG, and question-answering tasks. For named entity recognition (NER), we use Uni-NER \cite{zhou2023universalner}, Llama-3 8B Instruct \cite{dubey2024llama3herdmodels}, and Gemma-2 9B Instruct \cite{gemmateam2024gemma2improvingopen}. We also use Llama-3 as a Q/A model for the long-context task.

\noindent\textbf{Translation.} Translation is a common use case for language models. However, business or bureaucratic emails containing sensitive information face major privacy concerns pertaining to leakage of sensitive information \cite{lyu2024paradigmshiftfuturemachine}. For this task, we employ an \LLM~for named entity recognition (NER) of sensitive tokens belonging to the types of ([\textit{Name}], [\textit{Age}] and [\textit{Money}]). We evaluate \name's performance on 50 English-French and English-German samples obtained from WMT-14 \cite{bojar-etal-2014-findings} dataset. These samples are single sentences containing one or two PII values. We seek exact translations in this context and use BLEU scores as the quality oracle $\Q$ to assess their similarity to reference translations. We use \FPE~to sanitize [\textit{Name}] and [\textit{Money}], and use \textit{mLDP} for [\textit{Age}].

\noindent\textit{Results}. We report the BLEU scores for the translations of the original sentences and the ones obtained via \name~in Table~\ref{tab:en-de-small}. We observe that the BLEU scores are nearly identical in both cases, with only minor differences due to the performance variation of the translation model, nuances of language and NER. Details regarding the statistics of PII values, NER performance, details regarding encryption and ablations with larger privacy budgets can be found in App.~\ref{appendix:translation-results}. We present comparisons with Papillon~\cite{siyan2024papillon}, a contemporary privacy preserving framework in Section~\ref{sec:papillon}.

\noindent\textbf{Retrieval-Augmented Generation (RAG).} Retrieval-augmented generation is also commonly employed for a variety of LLM use cases \cite{gao2023retrieval}, including extraction of information from potentially sensitive documents. Typically, documents are split, embedded, and indexed in a vector database. At query time, the most relevant shards are retrieved based on lexical and semantic similarity, then provided as context to the \LLM~for answer generation. Our experiments focus on this final step: generating answers given a query and its relevant context. We consider two types of question-answering scenarios: numerical comparisons, and retrieval of factual information. We assess these settings by using GPT-4~\cite{openai2023gpt4} to generate tuples of \textit{Context} C, questions \textit{Questions} Q, and answers \textit{Answers} A; jointly sanitizing C and Q so that copies of the same sensitive attribute appearing in both C and Q are replaced with the same token, and comparing the desanitized LLM responses with A. Our numerical comparison questions involve comparing credit card balances and determining which is higher, while factual retrieval questions require returning specific aspects about a generated e-commerce order. The quality oracle $\Q$ is the accuracy of the answers.

\noindent\textit{Results.} We observe that \name~achieves $100\%$ accuracy for both the RAG tasks. Additional experimental details can be found in App.~\ref{appendix:RAG}

\noindent\textbf{Long-Context Q/A.} \LLM s can be tasked with not only retrieving specific information from long documents, but also integrating and reasoning about the information they contain. To simulate this, we use NarrativeQA~\cite{kocisky-etal-2018-narrativeqa}, a long-context, reading comprehension task. We answer questions about book or movie summaries, including character-related queries, using only the provided context. Character names are treated as sensitive attributes and sanitized using \FPE. Summaries have $534$ words on average with a standard deviation of $210$.

To assess the impact of \name~ on reasoning and reading comprehension, we use semantic textual similarity (STS) \cite{Reimers2019SentenceBERTSE} between the answers based on the original summaries and the answers based on \name~ summaries. This score also acts as the quality oracle $\Q$. We do not use BLEU scores for this experiment, as the reference answers only have a few words and do not capture any paraphrased response.  We discuss the performance of NER and present examples in App.~\ref{appendix:long-context-preempt}. We present comparisons with Papillon~\cite{siyan2024papillon}, a contemporary privacy preserving framework in Sec.~\ref{sec:papillon}.

\noindent\textit{Results.} We report the STS scores in Table~\ref{tab:long-context-small}. We find that \name~captures a significant amount of semantics of the plaintext response, with the GPT-4o response having an STS score of $0.934$. For context, if we don't desanitize the LLM response, the score drops to $0.523$ with respect to the plain responses. If the LLM gives a completely irrelevant response (such as the answer to an unrelated question), the score drops to around $0.146$. This demonstrates the robustness of the metric. Further details and ablations can be found in App~\ref{appendix:long-context-preempt}.

\begin{table}[t]
\centering
\caption{Semantic Textual Similarity scores of different methods for the Long-Context Q/A task. Higher value implies more similarity with the reference answer. ``Plain Responses'' refer to the responses for unsanitized inputs, and ``References'' indicate the ground truth responses. We find that \name~has a particularly high utility with respect to GPT-4o, outperforming prior methods. \name~uses Gemma-2 9B Instruct as the NER model for Gemini-1.5, and UniNER for Llama-3 and GPT-4o.}
\begin{tabular}{ l | c  c  c | c }
\hline
 \multirow{2}{*}{\textbf{STS Score}} & \multicolumn{3}{c|}{\textbf{\name}}  & \textbf{Papillon} \\ 
 & Llama-3 & Gemini-1.5 & GPT-4o & GPT-4o \\ 
 \hline \hline
 Plain Responses & 0.839 & 0.849 & 0.934 & 0.854 \\ 
 References (GT) & 0.514 & 0.722 & 0.510 & 0.458 \\ 
 \hline
\end{tabular}

\label{tab:long-context-small}
\end{table}
\noindent\textbf{Multi-Turn Financial Q/A.} LLMs are also frequently used in multi-turn conversational settings, and may be tasked with performing numerical reasoning over sensitive information. Thus, we assess \name~on a financial multi-turn question answering benchmark ConvFinQA \cite{chen2022convfinqaexploringchainnumerical}.The dataset consists of financial reports written by experts \cite{chen2022finqadatasetnumericalreasoning} followed by a sequence of  conversational, numerical-reasoning questions guiding the model through solving a multi-step problem. Each prompt includes background text and a table with yearly financial data spanning over several years. All numerical information (except years) is extracted using regex and sanitized using \mLDP. To handle repeated values within parentheses in the table, we use the helper string $\Psi$ in the regex updating of text to ensure that this structure is preserved in the sanitized text. ConvFinQA performance is typically reported in terms of exact match accuracy of responses. Since sanitization introduces noise in the numerical values, exact matching is no longer an appropriate evaluation criterion; instead, we measure utility after sanitization with the relative error of the prediction. Moreover, we observed that for this dataset, the answers returned by a model are occasionally correct up to the target answers sign or magnitude---often due to the questions being underspecified rather than model error. To account for the sensitivity of relative error to incorrect magnitudes of predictions, we check if the relative error of the magnitude and sign adjusted response is less than $.1$ of the correct answer. If the adjusted error is sufficiently small, we record it instead.

\noindent\textit{Results.} We report the 25th, 50th, and 75th percentiles of the relative error in GPT-4o's answers for sanitized and clean prompts. We further report the 25th, 50th, and 75th percentiles of ``consistency'', measured as the relative difference between the model's prediction on the sanitized query compared to the prediction on the clean query. We report the results in \autoref{tab:privacy-results}, observing a clear trend of performance improvement with a larger privacy budget, however, we note that at the 75th percentile, consistency does not change much. We also observe that the relative error of the sanitized prompts at the 75th percentile is lower than for unsanitized prompts, suggesting that addition of noise regularizes model behavior and prevents large outlier responses. We provide additional results with higher privacy budgets in App.~\ref{appendix:mutli-turn-finqa-rebuttal}.

\begin{table}[]
\centering
\caption{Performance evaluation on ConvFinQA benchmark with varying degrees of prompt sanitization ($\epsilon$ represents the privacy parameter for mLDP). Higher relative error indicates larger deviation from ground truth, while lower prediction consistency indicates a low relative discrepancy between sanitized and unsanitized responses. ``Base'' here indicates the baseline.}
\setlength{\tabcolsep}{3pt} 
\small 
\begin{tabular}{ l | c c c | c c c}
\hline
 \multicolumn{7}{c}{\textbf{Impact of \name~on ConvFinQA Performance}} \\
\hline
 \multirow{2}{*}{\textbf{$\epsilon$}} & \multicolumn{3}{|c|}{\textbf{Relative Error}} & \multicolumn{3}{|c}{\textbf{Prediction Consistency}} \\
 & 25th & Median & 75th & 25th & Median & 75th \\
\hline
\hline
0.1 & 0.0581 & 0.4000 & 4.4115 & 0.0698 & 0.3661 & 0.9994 \\
0.5 & 0.0154 & 0.0776 & 1.0000 & 0.0167 & 0.1345 & 0.9898 \\
1.0 & 0.0075 & 0.0408 & 0.9881 & 0.0084 & 0.0736 & 0.9899 \\
2.0 & 0.0040 & 0.0244 & 0.8686 & 0.0044 & 0.0447 & 0.9899 \\
\hline
Base & 0.0000 & 0.0000 & 12.6749 & - & - & - \\
\hline
\end{tabular}

\label{tab:privacy-results}
\end{table}

\subsection{Comparison of \texorpdfstring{\name}~ with Prior Methods} 
\label{sec:papillon}
We compare \name~with Papillon~\cite{siyan2024papillon}, a contemporary privacy-preserving framework. It uses a local LLM to create a proxy of the user query that omits all PII values, while attaining high utility with respect to a remote, task-performing model. We consider two tasks in our setting: \textit{Translation} and \textit{Long-Context Q/A}. We use GPT-4o for all steps of Papillon, with Llama-3.1 8B Instruct~\cite{dubey2024llama3herdmodels} as the local model.
\begin{table}[t]
\centering
\caption{BLEU scores for the English$\rightarrow$German and English$\rightarrow$French translation tasks, with UniNER-7B-PII for NER. All scores are w.r.t the reference translations from WMT-14. Higher value implies more similarity with the reference translation. We find that \name~has nearly identical performance with the translations of unmodified sentences and also outperforms prior methods.}
\begin{tabular}{ l | c  c | c  c  c}
\hline
 \multicolumn{6}{c}{\textbf{English $\rightarrow$ German, NER: UniNER-7B-PII}} \\
\hline
 \multirow{2}{*}{\textbf{Attribute}} & \multicolumn{2}{|c|}{\textbf{Gemini-1.5}} & \multicolumn{3}{|c}{\textbf{GPT-4o}} \\
 & Plain & \name & Plain & \name & Papillon\\
 \hline
 \hline
Name & 0.334 & 0.341 & 0.287 & 0.278 & 0.175 \\
Age & 0.235 & 0.252 & 0.243 & 0.231 & 0.135 \\
Money & 0.245 & 0.274 & 0.217 & 0.200 & 0.153\\
\hline
 \multicolumn{6}{c}{\textbf{English $\rightarrow$ French, NER: UniNER-7B-PII}} \\
 \hline
 \hline
Name & 0.423 & 0.408 & 0.432 & 0.419 & 0.290 \\
Age & 0.486 & 0.490 & 0.480 & 0.479 & 0.409\\
Money & 0.329 & 0.333 & 0.294 & 0.279 & 0.299 \\
\hline
\end{tabular}

\label{tab:en-de-small}
\end{table}

\begin{table*}[t]
\centering
\caption{Named entity recognition (NER) F1 scores for English (E), German (G), and French (F). Our finetuned version of UniNER either matches or outperforms all other models on almost every sensitive attribute. ``CCN'' and ``PN'' stand for \textit{Credit Card Number} and \textit{Phone Number} respectively.}
\label{tab:ner-pii-comparison}
\resizebox{\textwidth}{!}{\begin{tabular}{l|ccc|ccc|ccc|ccc|ccc|ccc}
\toprule
\multirow{3}{*}{\textbf{Attribute}} & \multicolumn{9}{c|}{\textbf{Part A: Open-source Models}} & \multicolumn{9}{c}{\textbf{Part B: Closed-source Models}}\\
\cmidrule{2-19}
& \multicolumn{3}{c|}{\textbf{Uni-NER-7B-PII}} & \multicolumn{3}{c|}{\textbf{Gemma-2 9B Inst}} & \multicolumn{3}{c|}{\textbf{Llama-3.1 8B Inst}} & \multicolumn{3}{c|}{\textbf{GPT-4.1}} & \multicolumn{3}{c|}{\textbf{Gemini-2.5}} & \multicolumn{3}{c}{\textbf{Claude 4 Sonnet}}\\
 & E & G & F & E & G & F & E & G & F & E & G & F & E & G & F & E & G & F\\
\midrule
Name & 1.00 & 1.00 & 1.00 & .907 & .893 & .846 & .836 & .766 & .715 & .843 & .883 & .845 & .742 & .903 & .840 & .791 & .867 & .872\\
Age & 1.00 & 1.00 & 1.00 & 1.00 & .951 & .990 & .960 & .884 & .822 & .970 & 1.00 & .990 & .990 & .990 & .990 & .980 & 1.00 & .990\\
Money & .940 & .860 & .880 & .940 & .827 & .824 & .820 & .710 & .820 & .882 & .941 & .959 & .990 & 1.00 & 1.00 & .990 & .980 & 1.00\\
SSN & .990 & 1.00 & .990 & .640 & .760 & .653 & .843 & .871 & .827 & .875 & .959 & .960 & .990 & 1.00 & 1.00 & .969 & .969 & .929\\
CCN & .980 & .960 & 1.00 & .952 & .962 & .873 & .916 & .926 & .855 & .971 & .971 & .980 & .980 & .990 & .970 & .980 & .980 & .980\\
Zipcode & 1.00 & .990 & .980 & .980 & .990 & .980 & .952 & .925 & .936 & .980 & .962 & .981 & .990 & .980 & .990 & .990 & .942 & .990\\
Date & 1.00 & 1.00 & 1.00 & .778 & .708 & .649 & .960 & .846 & .895 & 1.00 & .980 & .970 & .860 & .733 & .784 & 1.00 & .971 & .980\\
Password & .980 & 1.00 & 1.00 & .885 & .887 & .833 & .238 & .087 & .163 & 1.00 & .970 & .950 & .980 & .810 & .970 & .990 & .970 & .922\\
Sex & 1.00 & 1.00 & 1.00 & .971 & .943 & .980 & .926 & .673 & .830 & .962 & .945 & .971 & .971 & .925 & .990 & .971 & .954 & .980\\
PN & .980 & 1.00 & 1.00 & .952 & .971 & .962 & .926 & .971 & .971 & .980 & .970 & .990 & .990 & 1.00 & .990 & .980 & .980 & .990\\
\bottomrule
\end{tabular}}
\end{table*}
\noindent\textbf{Translation.} We consider 100 samples for each sensitive attribute ([\textit{Name}],[\textit{Age}],[\textit{Money}]) for both languages, for a total of 600 samples. Following Papillon, we create optimized prompts for each attribute-language pair.

\noindent\textit{Results.} We report BLEU scores in Table~\ref{tab:en-de-small}. We find that \name~significantly outperforms Papillon, except for the [\textit{Age}] and [\textit{Money}] PII categories for the English-French translation task, where it is comparable. Furthermore, the average leakage of privacy due to NER failure is $71\%$ of unique PII values compared to $97\%$ for \name. We detail their relative performance in App.~\ref{appendix:translation-papillon}.

\noindent\textbf{Long-Context Q/A.} We consider 50 samples for prompt optimization. Each sample contains a unique summary, a question based on it, and the corresponding answer.

\noindent\textit{Results.} We report the STS scores in Table~\ref{tab:long-context-small}. We find that \name~performs somewhat better than Papillon. However, Papillon as implemented, is mostly unsuccessful in preventing leakage for long context tasks. We observed that $80\%$ of all prompts passed to the remote model contain character identities. These prompts are just the questions and do not include the summary. As the summaries are based off Wikipedia entries, the remote model is able to identify those characters and correctly respond to the query. We discuss NER performance and examples of successes and failures in App.~\ref{appendix:long-context-papillon}. 
\subsection{Impact of Design Choices on \texorpdfstring{\name}~ Utility}
\label{exp:design-choices}
To address \textbf{Q3}, we examine design choices for two components: NER and encryption format.
\noindent\textbf{Named-Entity Recognizer (NER).} Uni-NER~\cite{zhou2023universalner} is an \LLM~trained for generic named entity recognition. We finetune it on $10$ high-risk categories from the AI4Privacy dataset~\cite{ai4privacy_2023}. We evaluate the NER as a type annotator on a held out subset of the dataset, consisting of text in English, German and French, with $50$ samples per category, per language.  We tabulate results for the following categories: ``Money'', ``Name'', ``Age'', ``SSN'', ``Credit Card Number'', ``Zipcode'', ``Date'', ``Password'', ``Sex'', ``Phone Number''. We make comparisons with off-the-shelf proprietary and open-source models, including: GPT-4.1~\cite{gpt4.1}, Claude 4 Sonnet~\cite{claude}, Gemini 2.5~\cite{gemini2.5}, Llama-3.1 8B Instruct~\cite{dubey2024llama3herdmodels} and Gemma-2 9B Instruct~\cite{gemmateam2024gemma2improvingopen} (details in App.~\ref{app:ner-enc}). As our experiments only deal with \textit{Name}, \textit{Age} and \textit{Money}, we use another Uni-NER model, specifically finetuned on these attributes.

\noindent\textit{Results.} As seen in Table~\ref{tab:ner-pii-comparison} (App.~\ref{app:ner-enc}), fine-tuning Uni-NER on these high-risk categories yields 100\% F-1 scores across most attributes, often exceeding the performance of state-of-the-art proprietary models. \begin{table*}[tbh!]
\caption{A comparison of prompt sanitization frameworks as per the design goals in Sec.~\ref{sec:name:design}. We find that \name~is the only framework that has all the desirable qualities of a secure and high utility prompt sanitizer.}
    \centering
    \scalebox{0.95}{\begin{tabular}{lcccc}
    \toprule
    Method & Stateless & Formal Privacy Guarantee & High Utility & Resource-Efficient\\
    \midrule
        \textbf{\name} (Ours) & \cmark & \cmark & \cmark & \cmark\\
        Papillon~\cite{siyan2024papillon} & \xmark & \xmark & \cmark & \cmark\\
        Substitution-based~\cite{shen2024thiefkeeperbalancingusability, kan2023protecting,chen2023hide} & \xmark & \xmark & \cmark &  \cmark\\
        Cryptography-based~\cite{cryptoeprint:2023/1147, hao2022iron,chen2022x} & \cmark & \cmark & \cmark &  \xmark\\
        DP-based noising of text~\cite{Duan2024,tang2023privacy,hong2023dp,tian2022seqpate} & \cmark  & \cmark & \xmark & \xmark\\
        DP-based noising of text embeddings~\cite{feyisetan2020privacy} & \cmark & \cmark & \xmark  & \cmark\\
        DP-based noising of tokens~\cite{Carvalho2021TEMHU, arnold2023driving,arnold2023guiding,chen2023customized}& \cmark &  \cmark & \xmark & \cmark\\
        DP-based text paraphrasing~\cite{mattern2022limits, utpala2023locally, lewis2020bart, igamberdiev2023dpbart} & \cmark & \cmark & \xmark &  \cmark\\
    \bottomrule
    \end{tabular}}
    \label{tab:related-work-comparison}
\end{table*}

\noindent\textbf{Encryption Format.} \name~assumes that the \LLM's performance depends on preserving the format of type $\t_{\texttt{I}}$ tokens. We validate this assumption by evaluating the \LLM~on two other sanitization algorithms: $(1)$ that does not preserve the format at all, and $(2)$ that preserves an incorrect format. For the first case, we sanitize the type $\t_{\texttt{I}}$ tokens with AES, which replaces the sensitive tokens with 16 bytes of random strings. In the second case, we randomly substitute the tokens without maintaining the correct format (e.g. replacing a 5-digit ZIP code with a randomly chosen 8-digit value). We assess this in the context of a RAG task by generating 31 tuples of contexts (C), questions (Q), and answers (A) corresponding to a factual retrieval task. For each tuple, we evaluate the percentage of correct, desanitized answers using GPT-4.

\noindent\textit{Results.} We observe that our model achieves 100\% accuracy in factual information retrieval when employing \FPE. However, performance drops to 70.97\% with AES encryption and 77.42\% with random substitution using incorrect formats. This confirms that format preservation is crucial for the \LLM's performance. We provide additional results for the Translation and Multi-Turn Financial Q/A tasks with different privacy budgets in App.~\ref{appendix:translation-results} and App.~\ref{appendix:mutli-turn-finqa-rebuttal} respectively.

%% file: NDSS_revision/5_Discussion.tex
\section{Related Work}\label{sec:relatedwork}

A line of work proposes to sanitize the prompts via substitution using a local \LLM
\cite{shen2024thiefkeeperbalancingusability, kan2023protecting,chen2023hide}. However, such solutions cannot be stateless if they intend to provide utility by desanitizing \LLM~responses. Cryptographic methods have also been explored for  protecting user privacy at inference~\cite{cryptoeprint:2023/1147, hao2022iron,chen2022x}. However, these approaches impose high computational and communication overheads. 
One line of approach for protecting privacy at inference involves employing DP for in-context learning by generating a synthetic dataset~\cite{Duan2024,tang2023privacy,hong2023dp,tian2022seqpate}. However, these approaches are only applicable when a large collection of data is available, and are different from sanitizing an individuals sensitive information when they are submitting a simple query to an \LLM. More similar to our setting are local DP based approaches. However, a key difference from our work is the way in which noise is added. A line of prior work employs metric DP by adding noise to text embeddings, and then decoding the private embeddings back into text~\cite{feyisetan2020privacy}: this violates the definition of a prompt sanitizer as this might not preserve the types of the tokens (Sec. \ref{sec:scheme}). Another approach noisily sample a token from a pre-defined list of ``similar'' tokens~\cite{Carvalho2021TEMHU, arnold2023driving,arnold2023guiding,chen2023customized} which require carefully selecting the list of similar tokens. 
Another line of work generates a noisy paraphrase of the prompts~\cite{mattern2022limits, utpala2023locally, lewis2020bart, igamberdiev2023dpbart}. However, these methods suffer from the curse of dimensionality as the amount of noise grows proportionally with the length of the generated text leading to poor utility. Table~\ref{tab:related-work-comparison} provides a summary comparing \name~with prior work.

%% file: NDSS_revision/7_Conclusion.tex
\section{Conclusion}\label{sec:conclusion}
\LLM s introduce new challenges for protecting sensitive information at inference time. We address this by introducing a cryptographically inspired primitive—the prompt sanitizer—which transforms prompts to protect sensitive tokens. We then present \name, a system that implements this primitive with provable privacy guarantees. Experiments show that \name~maintains high utility across both structured and open-ended prompts.

%% file: NDSS_revision/Appendix_ndss_revision.tex
\clearpage
\appendix
\subsection{Background on \FPE}\label{app:FPE}
\noindent\textbf{Mechanism for $\epsilon$-\mLDP.}
\label{app:emldp-algo}
\begin{algorithm}
 \caption{Mechanism $\Mdp$}\label{alg:1}
\begin{algorithmic}[1]
\Statex   \textbf{Input:} $x$ - Plaintext; $\epsilon$ - Privacy parameter; $[k]$ - Output domain
\Statex \textbf{Output:} $o'$ - Noisy encoding;
\State \textbf{for} $x \in \mathcal{X}$:
\State \hspace{0.3cm}\textbf{for} $i \in [k]$ 
\State
\begin{equation}
\hspace{0.5cm}
p_{x,i} = \frac{e^{-|x-i|\cdot \epsilon/2}}{\overset{k}{\underset{j=1}{\sum}}
e^{-|x-j|\cdot \epsilon/2}} \hspace{0.3cm}
\end{equation}
\State \hspace{0.3cm}\textbf{end for}
\State \hspace{0.3cm} $p_x=\{p_{x,1},\cdots, p_{x,k}\}$  
\State \textbf{end for}
\State $o\sim p_x$ 
\State \textbf{return} $o $ 
\end{algorithmic}\label{alg:Mdp}
\end{algorithm}

\noindent\textbf{Security Definition of \FPEs.}
Pseudo-Random Permutation (\PRP)
security requires that an adversary cannot distinguish encryptions with a randomly chosen key from random permutations
over the format domain; single-point indistinguishability (\SPI)
requires that the adversary cannot distinguish the encryption of
any message of its choice from a random ciphertext; message
privacy (\MP) requires that ciphertexts reveal no information
on the encrypted message, except its format; and
similar to \MP, but weaker than it, message recovery (\MR)
only requires that the ciphertext does not completely reveal the
encrypted message. Bellare et
al.~\cite{BRRS2009} show that 
\begin{gather}
    \PRP \iff \SPI \Rightarrow \MP \Rightarrow \MR
\end{gather} This implies that  \PRP~is the strongest security notion and \MR~is the weakest. We
note that though \PRP~is the best security notion one can hope
to achieve for \FPEs, the three weaker notions can, in many
concrete cases offer much better efficiency and may therefore suffice in practice. Most of the schemes in practice focus on \MP~or \MR~security guarantees.
\subsection {Proof of Thm. \ref{thm:mLDP}} \label{app:thm}

\begin{proof}

For all $x\in \mathcal{X}$ and $i \in [k]$, we have
\begin{gather*}\hspace{-4cm}\frac{\mathrm{Pr}\big[\Mdp(x,\epsilon)=i]}{\mathrm{Pr}\big[\Mdp(x+t,\epsilon)=i\big]}= \\ \hspace{2cm}\Big(e^{(|x+t-i|-|x-i|)\cdot \epsilon/2} \cdot\frac{\overset{k}{\underset{j=1}{\sum}}e^{-|x+t-j|\cdot \epsilon/2}} {\overset{k}{\underset{j=1}{\sum}}e^{-|x-j|\cdot \epsilon/2}} \Big)\\\leq e^{t\epsilon/2}\cdot e^{t\epsilon/2} \\\hspace{0.2cm}\big[\because  |x-j|-t\leq|x+t-j|\leq |x-j|+t\big] \\= e^{t\epsilon }  \label{eq:DP2} \numberthis\end{gather*}
Similarly, \begin{gather*}\frac{\mathrm{Pr}\big[\Mdp(x,\epsilon)=i]}{\mathrm{Pr}\big[\Mdp(x+t,\epsilon)=i\big]}\geq e^{-t\epsilon}
\end{gather*}

\end{proof}

\subsection{Experimental Setup Details}
\label{appendix:experiment-setup}
\subsubsection{Datasets}
\begin{itemize}
    \item Translation: Subsets of WMT-14 dataset for offline translation \footnote{\url{https://huggingface.co/Helsinki-NLP/opus-mt-de-en} for German and \url{https://huggingface.co/Helsinki-NLP/opus-mt-fr-en}}.
    \item RAG: Simple GPT4-generated datasets for numerical comparisons and factual information retrieval.
    \item Long-Context Q/A: Subsets of the NarrativeQA dataset.
    \item Multi-Turn FinancialQ/A: We use the ConvFinQA dataset.
    \item Named Entity Recognition: For fine-tuning NER model we use the AI4Privacy dataset (referred to as the PII dataset).
\end{itemize}

\subsubsection{Frameworks and Infrastructure}
\begin{itemize}
    \item We use an implementation (pyfpe) of the FF3 algorithm \cite{Dworkin201380038GR} for format preserving encryption and a custom mLDP sanitization mechanism available in the provided code.
    \item GPT-4o \cite{openai2023gpt4}, Gemini-1.5 \cite{geminiteam2024gemini15unlockingmultimodal}, and OPUS-MT~\cite{opus} for translations, RAG, and question-answering tasks.
    \item For NER we use Uni-NER \cite{zhou2023universalner}, Llama-3 8B Instruct \cite{dubey2024llama3herdmodels}, and Gemma-2 9B Instruct \cite{gemmateam2024gemma2improvingopen} as open source models and GPT-4o\cite{openai2023gpt4} and Gemini 1.5\cite{geminiteam2024gemini15unlockingmultimodal} as closed source models.
    \item 3 A100 GPUs for NER model fine-tuning, with the following configuration: half-precision (bfloat16), 16 gradient accumulation steps, learning rate of 2e-5, weight decay of 0, cosine learning rate schedule. 
\end{itemize}

\subsection{Impact on Utility (Q1)}

\subsubsection{Translation Task Details}
\label{appendix:translation}
For translation tasks we employ an LLM for NER of sensitive tokens belonging to the sensitive types of ([\textit{Name}], [\textit{Age}] and [\textit{Money}]).\\ 




\noindent\textbf{Name:} Names are sanitized using \FPE. We curate $1000$ European first and last names using the First and Last Names Database~\cite{NameDataset2021}. Then for any full name found during NER, we have two cases: (1) the first and last names exist in our curated list or (2) they don't, so we replace the last indices of each list with our found first and last names. Now, each name will have a certain index on the list between $000-999$. Using both of these, we get a six digit number representing our found name. We can now apply standard FPE to obtain a new six digit number. The first three digits correspond to the index of the encrypted first name, and the last three digits correspond to the index of the encrypted last name. In case the found name is only a first (or last) name, we choose a default last (or first) name.\\

\noindent\textbf{Money:} Money is also sanitized using \FPE. We simply add six ``$9$''s to the front of the found monetary value and apply FPE. This also preserves any commas or spaces in the number. However, this does lead to some bizarre looking values such as ``$6.07$'' being encrypted as ``$7728491.89$'', which makes it difficult for NER to pick up.\\

\noindent\textbf{Age:} Age is sanitized over the domain of two digit numbers in integer increments$[10,11,\dots,98,99]$; both settings employ an $\epsilon = 1$. The sanitized prompt is submitted to an LLM for translation, and the NER model is again used to annotate the output text for desanitization.\\

To demonstrate the robustness of~\name, we use two modes of translation:    
\begin{itemize}
    \item Online models such as Gemini-1.5~\cite{geminiteam2024gemini15unlockingmultimodal} and GPT-4o~\cite{openai2023gpt4}
    \item Dedicated offline models like OPUS-MT \cite{opus}
\end{itemize} 

To evaluate impact of sanitization on performance, we sample 50 English to German and English to French string pairs from the WMT-14 \cite{bojar-etal-2014-findings} dataset. As multiple translations can be valid and outputs of frontier models are not always deterministic, instead of testing for exact matches between desanitized translations of sanitized text translations of the original text, we compute BLEU scores to assess their respective similarities to reference translations. \\

\subsubsection{Translation Results}
\label{appendix:translation-results}
\begin{figure}[t]
    \centering
    \includegraphics[width=\linewidth]{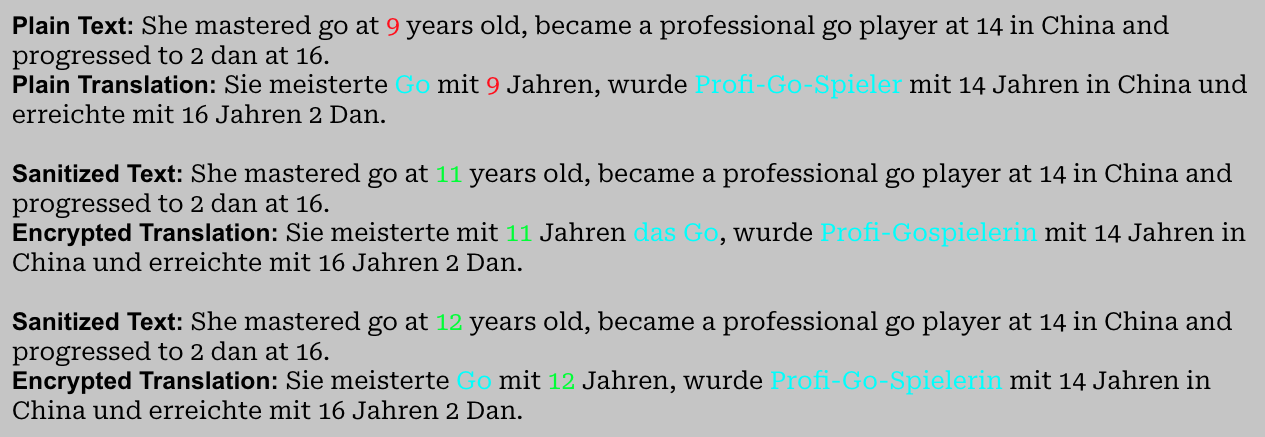}
    \caption{Sanitized values can affect the grammatical structure of sentences. When we change a sensitive value like age from `9' in the plain text (highlighted in red), to `11' in the sanitized text, the output translations also change (differences highlighted in blue). However, this occurs only for the number `11' and returns to the original sentence structure if it is replaced with `12' or any other value less than 100.Translation done using OPUS-MT, from English to German. }
    \label{fig:ap_ablation-1}
\end{figure}

\begin{figure}[t]
    \centering
    \includegraphics[width=\linewidth]{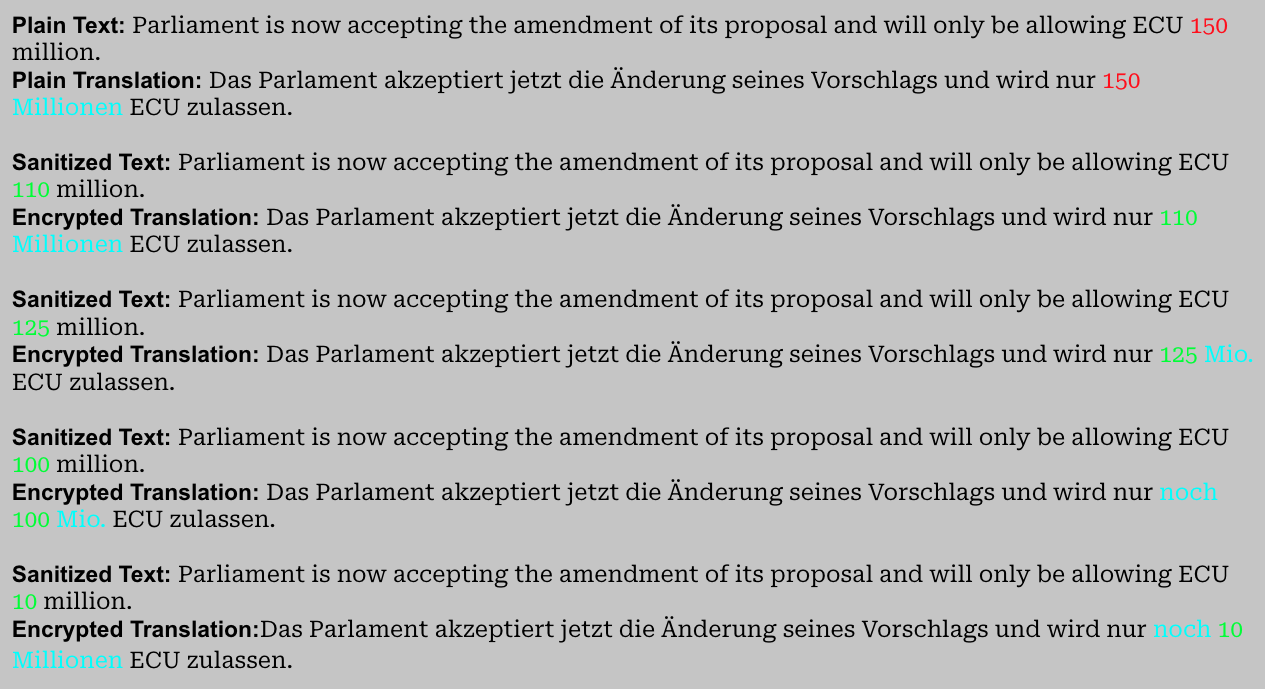}
    \caption{Changes in output translations for different sanitized values. As we see, the numerical money value (green) can change the words that come before and after it (blue). Translation done using OPUS-MT, from English to German.}
    \label{fig:ap_ablation-2}
\end{figure}

We report the BLEU scores of translated plain text and \name~text (after desanitizing). We find that translation is largely invariant to prompt sanitization. This is true across different NER and translation models and different languages. Table~\ref{tab:en-de} and Table~\ref{tab:en-fr} show results with different models for NER and a additional model for translation. We make the following observations:

\begin{enumerate}
    \item We find there is only a marginal difference in the quality of plain and \name~translations across all types. We note that GPT-4o and Gemini-1.5 do sampling during text generation, thereby making their outputs non-deterministic. However, the plain and \name~translations are of high quality.
    \item We found a significant number of translations that are identical. However among the mismatched samples, we observed that translated sentence structures can vary due to the value of sanitized text, as shown in Figure \ref{fig:ap_ablation-1} and Figure~\ref{fig:ap_ablation-2}. 
    \item We also note that the NER model would occasionally misidentify parts of the sensitive attribute. For example, if a monetary value was $121.445$, it might only identify $121.4$ as the sensitive attribute. This hinders performance during desanitization. However, we believe this can be solved with better finetuning and prompt engineering.
\end{enumerate}

Language artifacts like these can make sanitation for exact translation difficult in practice. We leave a detailed analysis of this phenomena for future work. 

We also note that Gemma-2 and Llama-3 as NER models don't perfectly catch all PII values in the test samples. These samples are ignored while calculating BLEU scores. Specifically, the privacy leakage in terms of unique PII values missed are: 
\begin{itemize}
    \item \textbf{Gemma-2:} $4.8\%$ on average, for both tasks.
    \item \textbf{Llama-3:} $15.4\%$ on average, for both tasks.
\end{itemize}

\begin{figure}
    \centering
    \begin{subfigure}[b]{0.49\textwidth}
         \centering
         \includegraphics[width=\textwidth]{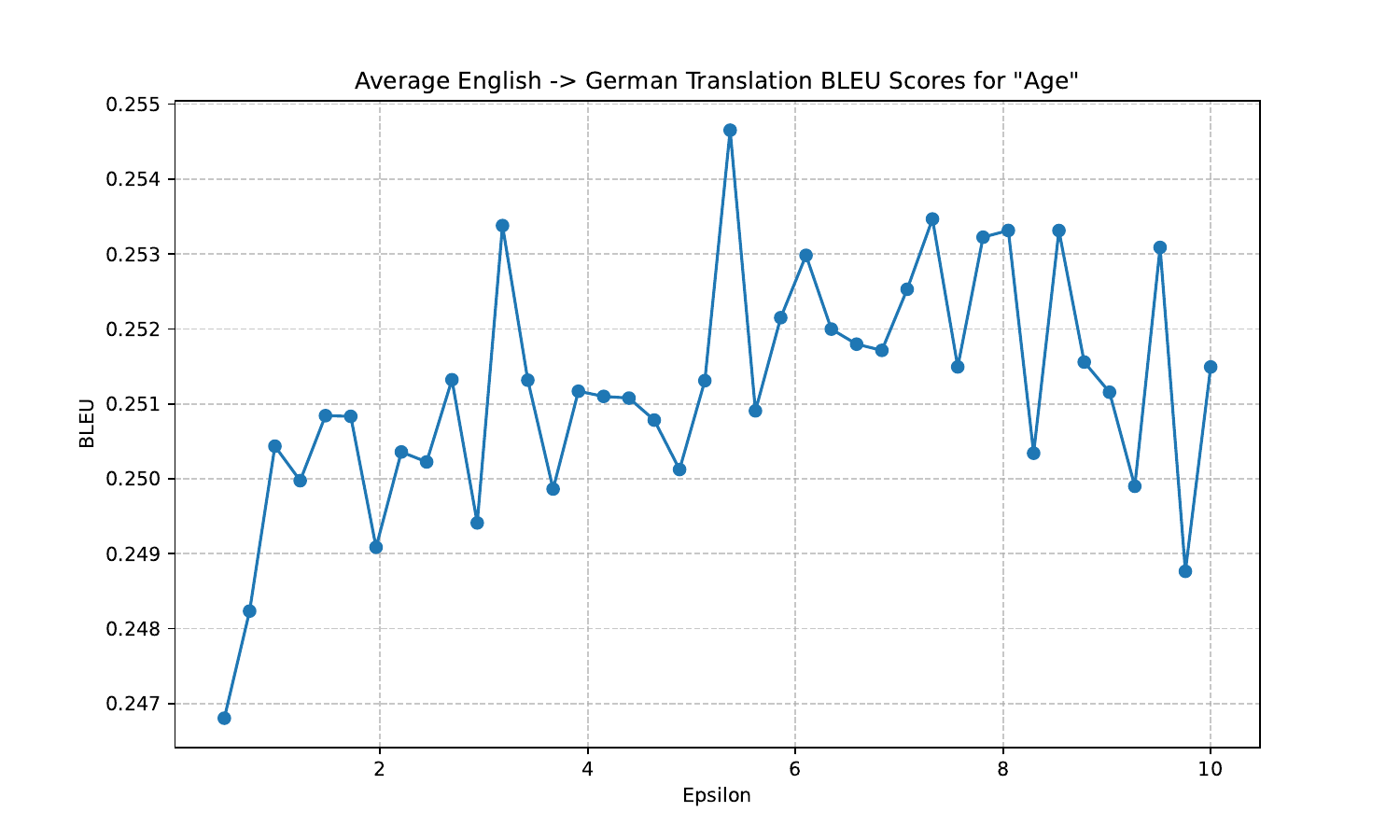}
         \caption{Translation: English $\rightarrow$ German}
         \label{fig:translation_age_tradeoff_de}
     \end{subfigure}
     \begin{subfigure}[b]{0.49\textwidth}
         \centering
         \includegraphics[width=\textwidth]{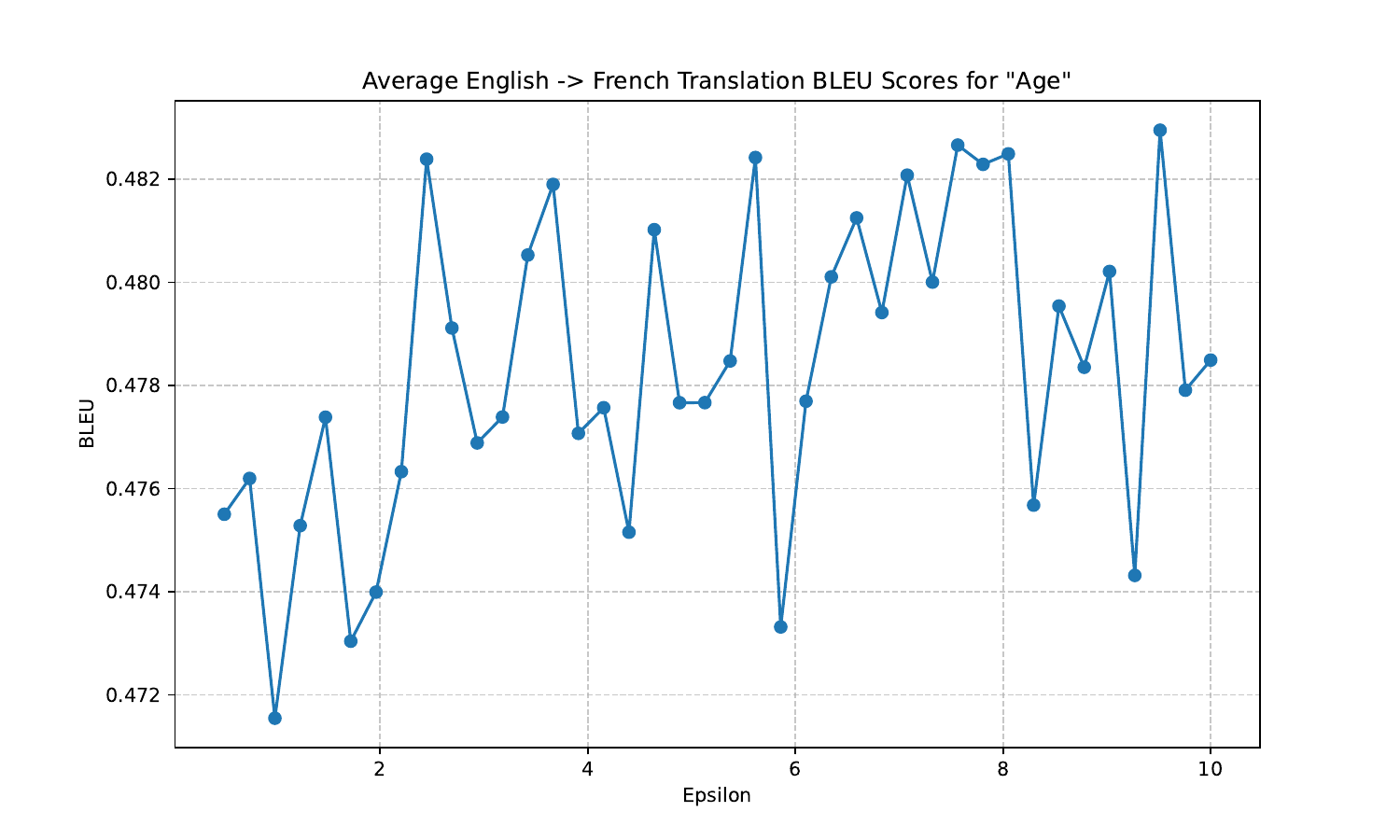}
         \caption{Translation: English $\rightarrow$ French}
         \label{fig:translation_age_tradeoff_fr}
     \end{subfigure}
    \caption{Average BLEU scores for translation where the PII is `Age', with increasing privacy budgets. We found that utility improves with larger values of $\epsilon$.}
    \label{fig:translation_age_tradeoff}
\end{figure}

\noindent\textbf{Additional Results at Higher Privacy Budgets:} We present additional results for the translation task when considering `Age' as the PII attribute in Fig.~\ref{fig:translation_age_tradeoff}. We found a weak, but upward trend in BLEU scores with increasing values of $\epsilon$ (0.5 to 10), averaged across three different translation models and three seeds each. We believe that the signal is weak because the efficacy of translation is not critically dependent on the numerical value of the sensitive token considered here (‘Age’).\\

\noindent\textbf{Performance of NER:} NER performs better for English than for French and German. This impacts \name's utility, as desanitization involves running NER on the LLM's response (translations in French and German). \\

\noindent\textbf{Distribution of PII values:} Since we want to measure the impact of sanitization on utility, we look up sanitized PII values for [\textit{Name}] and [\textit{Money}], instead of using NER in the desanitization phase. We found that $96\%$ of all unique PII values were identified and sanitized by \name, when using UniNER as the NER model. There are around $1.3$ unique PII values per sentence on average, with around $1$ instance per PII value in each sentence for both languages. More details regarding encryption and ablations can be found in App.~\ref{appendix:translation} and App.~\ref{appendix:translation-results}. \\

\begin{figure}
    \centering
    \includegraphics[width=\linewidth]{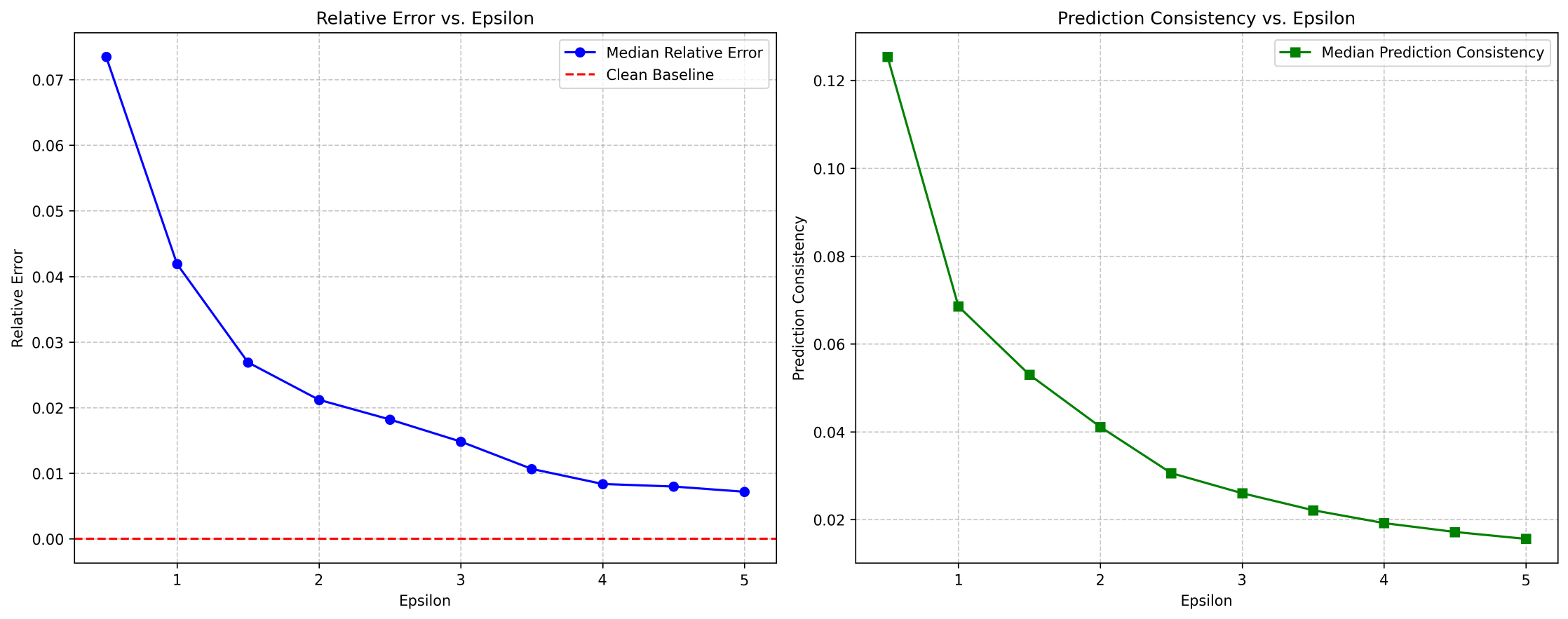}
    \caption{Median relative error for the Multi-turn Financial Q/A task, with increasing privacy budgets. We observe a consistent and smooth trend of median relative error (capturing utility of the model responses), improving with higher epsilon}
    \label{fig:multi_turn_rebuttal}
\end{figure}

\begin{table*}[t]
\centering
\caption{BLEU scores for the English$\rightarrow$German translation task, with different NER models. All scores are with respect to the reference translations from WMT-14.}
\begin{tabular}{ l | c  c | c  c | c  c }

\hline
\multicolumn{7}{c}{\textbf{Part A: UniNER-7B-PII}} \\
\hline
 \multirow{2}{*}{\textbf{Attribute}} & \multicolumn{2}{|c|}{\textbf{Gemini-1.5}} & \multicolumn{2}{|c|}{\textbf{GPT-4o}} & \multicolumn{2}{|c}{\textbf{OPUS-MT}} \\
 & Plain & \name & Plain & \name & Plain & \name \\
 \hline
Name & 0.334 & 0.341 & 0.287 & 0.278 & 0.319 & 0.306 \\
Age & 0.235 & 0.252 & 0.243 & 0.231 & 0.294 & 0.282 \\
Money & 0.245 & 0.279 & 0.217 & 0.200 & 0.238 & 0.225 \\
\hline
\multicolumn{7}{c}{\textbf{Part B: Llama-3 8B Instruct}} \\
\hline
Name & 0.338 & 0.304 & 0.292 & 0.265 & 0.319 & 0.297 \\
Age & 0.273 & 0.248 & 0.242 & 0.231 & 0.310 & 0.298 \\
Money & 0.256 & 0.269 & 0.219 & 0.193 & 0.238 & 0.228 \\
\hline
\multicolumn{7}{c}{\textbf{Part C: Gemma-2 9B Instruct}} \\
\hline
Name & 0.336 & 0.307 & 0.292 & 0.258 & 0.319 & 0.303 \\
Age & 0.262 & 0.225 & 0.238 & 0.218 & 0.297 & 0.286 \\
Money & 0.260 & 0.243 & 0.219 & 0.193 & 0.238 & 0.223 \\
\hline
\end{tabular}

\label{tab:en-de}
\end{table*}

\begin{table*}[t]
\centering
\caption{BLEU scores for the English$\rightarrow$French translation task, with different NER models. All scores are with respect to the reference translations from WMT-14.}
\begin{tabular}{ l | c  c | c  c | c  c }
\hline
\multicolumn{7}{c}{\textbf{Part A: UniNER-7B-PII}} \\
\hline
\multirow{2}{*}{\textbf{Attribute}} & \multicolumn{2}{|c|}{\textbf{Gemini-1.5}} & \multicolumn{2}{|c|}{\textbf{GPT-4o}} & \multicolumn{2}{|c}{\textbf{OPUS-MT}} \\
& Plain & \name & Plain & \name & Plain & \name \\
 \hline
Name & 0.423 & 0.403 & 0.432 & 0.419 & 0.415 & 0.412 \\
Age & 0.486 & 0.490 & 0.480 & 0.479 & 0.470 & 0.471 \\
Money & 0.329 & 0.333 & 0.294 & 0.279 & 0.380 & 0.367 \\
\hline
\multicolumn{7}{c}{\textbf{Part B: Llama-3 8B Instruct}} \\
\hline
Name & 0.428 & 0.365 & 0.432 & 0.365 & 0.415 & 0.373 \\
Age & 0.480 & 0.487 & 0.473 & 0.471 & 0.470 & 0.471 \\
Money & 0.353 & 0.353 & 0.297 & 0.282 & 0.380 & 0.359 \\
\hline
\multicolumn{7}{c}{\textbf{Part C: Gemma-2 9B Instruct}} \\
\hline
Name & 0.407 & 0.381 & 0.430 & 0.361 & 0.415 & 0.363 \\
Age & 0.497 & 0.457 & 0.491 & 0.489 & 0.474 & 0.474 \\
Money & 0.330 & 0.330 & 0.307 & 0.282 & 0.380 & 0.360 \\
\hline
\end{tabular}

\label{tab:en-fr}
\end{table*}

\subsubsection{Retrieval-Augmented Generation (RAG) Task Details}
\label{appendix:RAG}
The aim of these experiments is to investigate if the sanitization of sensitive tokens of the following types -- [\textit{Money}], [\textit{CCN}], [\textit{SSN}], [\textit{ZipCode}] and [\textit{Name}] --- impacts the correctness, and therefore utility of the LLM responses, in two types of question-answering scenarios: numerical comparisons, and retrieval of factual information.\\

\noindent\textbf{Numerical Comparisons: } We generate 20 tuples of \textit{Context}, \textit{Questions} and \textit{Answers} using GPT4, where:
\begin{enumerate}
    \item Context C is a few sentences describing financial details (jobs, salaries, credit debt, etc), which contain sensitive items like social-security numbers (SSN), credit-card numbers (CCN), salaries, credit-card balance.
    \item Question Q is a binary-choice comparison question, e.g. ``Which credit card has higher balance?''
    \item Answer A is the correct answer, indicating choice 1 or 2. \footnote{we can equivalently set the correct answers to be the credit card numbers themselves, however this format is allows for a standard prompting template for all queries}
\end{enumerate}

For each such tuple C,Q, we apply \name~to sanitize the LLM input and desanitize the LLM response to compare with the correct answer A. We use GPT-4 to perform NER, FPE to sanitize [\textit{CCN}], [\textit{SSN}], [\textit{ZipCode}] and [\textit{Name}] and mLDP with $\epsilon=1$ to sanitize [\textit{money}] such as salaries and credit card balances. We sanitize C and Q jointly, i.e. copies of the same sensitive attribute across C and Q are updated using an LLM as well as verifying they are copies if they are both annotated as the same type and have the same value.\\

\textbf{Retrieval of factual information}
We use GPT-4 to generate an e-commerce question/answering dataset consisting of 30 tuples. Each tuple consists of a Context C, Question Q and Answer A where:
\begin{enumerate}
    \item Context is the description of customer orders, containing order IDs, cost, total cost with shipping, estimated arrival dates and shipping zip code.
    \item Question is a customer question about single aspect of their order (e.g. cost, estimated arrival etc)
    \item Answer is the correct response to the question, as a phrase.
\end{enumerate}
Our procedure for applying \name~is the same as for numerical comparison tasks, with correctness evaluated by GPT-4. \\

\subsubsection{RAG Results}
We observe that our model achieves $100\%$ accuracy for retrieval of factual information tasks and comparison tasks when using FPE as our encryption method.\\

\subsubsection{Long-Context Q/A Task Details} 
\label{appendix:long-context-preempt}
We sanitize names using FPE, with the same process described in App.~\ref{appendix:translation}.\\

\noindent\textbf{Failure Modes during NER:} There are an average of $8.04$ unique identities in each
summary, and \name is able to correctly identify and sanitize $92\%$ of all unique identities. There were instances where NER fails, particularly during the desanitization phase and our method's performance decreases as a result. We found that when the name is more exotic (for example, ``Co-tan'') NER will fail to pick it up. In fact, this extends to desanitization, where encrypted names tend to be exotic, such as ``Paradise Arnoldo'' and ``Gheorghe Adamaène" are missed by NER. If there are two consecutive names (typically when a character is introduced, followed by the actor's name in braces), NER will fail to pick up the second name. These are typically actor names mentioned only once and are not pertinent to the questions. Both models will occasionally misidentify titles such as ``Count'' or ``the King'' as valid names. They also occasionally misidentify pronouns (such as ``him'', ``she'') and articles (such as ``the'') which are filtered accordingly. Lastly, UniNER occasionally hallucinates common names such as ``Josh'', but these typically don't occur in the summaries, so performance is unaffected.\\

\noindent\textbf{Additional STS Scores for \name:} We report additional STS scores with respect to plain responses without desanitization (``Encrypted''), and irrelevant answers (``Irrelevant'') in Table~\ref{tab:long-context-big}. On average, we see that \name~is able to capture the semantics of the passage even with all PII values sanitized and maintains high utility with open and closed-source models. Without desanitization, the STS score falls dramatically for all Q/A models. If the answers are completely irrelevant then the scores drop to 0.15 on average, showing the robustness of the metric.\\

\begin{table}[t]
\centering
\caption{Semantic Textual Similarity scores of different methods for the Long-Context Q/A task. Higher value implies more similarity with the reference answer. ``Plain Responses'' refer to the responses for unsanitized inputs. ``References'' indicate the ground truth responses. ``Encrypted'' indicates the STS scores of \name~responses without desanitization, with respect to the plain responses. Lastly, ``Irrelevant'' refers to the STS scores with an irrelevant reference answer. We find that \name~has a particularly high utility with respect to GPT-4o, outperforming prior methods. \name~uses Gemma-2 9B Instruct as the NER model Gemini-1.5 and UniNER for Llama-3 and GPT-4o.}
\begin{tabular}{ l | c  c  c | c }
\hline
 \multirow{2}{*}{\textbf{STS Score}} & \multicolumn{3}{c|}{\textbf{\name}}  & \textbf{Papillon} \\ 
 & Llama-3 & Gemini-1.5 & GPT-4o & GPT-4o \\ 
 \hline \hline
 Plain Responses & 0.839 & 0.849 & 0.934 & 0.854 \\ 
 References (GT) & 0.514 & 0.722 & 0.510 & 0.458 \\ 
 Encrypted & 0.450 & 0.496 & 0.523 & - \\
 Irrelevant & 0.148 & 0.166 & 0.146 & 0.141\\
 \hline
\end{tabular}

\label{tab:long-context-big}
\end{table}

\subsubsection{Additional Results for Multi-Turn Financial Q/A} 
\label{appendix:mutli-turn-finqa-rebuttal}
We include additional results for this task when considering a broad range of privacy budgets in Fig.~\ref{fig:multi_turn_rebuttal}. We observe a consistent and smooth trend of median relative error (capturing utility of the model responses), improving with higher $\epsilon$ (0.5 to 5). The Financial Q/A dataset primarily consists of numerical reasoning and retrieval, making its utility much more sensitive to the amount of noise present; as long as the model’s underlying reasoning is correct, the amount of noise would have a direct functional relationship with the relative error, explaining the consistent and smooth improvements.
\\

\subsection{Comparison with Prior Methods (Q2)}

\subsubsection{Comparison with Strawman Solutions}
\label{appendix:strawman}

\noindent\textbf{Redaction:} For example, consider a factual information retrieval task in the context of RAG for financial documents, with the prompt: ``Please return all bank accounts with balance greater than \$2000.'' If the sensitive tokens of the types [\textit{Bank A/c}] or [\textit{Bank Balance}] are redacted, the \LLM~is unable to provide any useful response. Furthermore, any tasks which rely on operations over sensitive information, such as numerical reasoning, would similarly provide little to no utility, even if redacted tokens are replaced by a "best guess" of a masked language model.\\

\subsubsection{Translation Task - Papillon} 

\label{appendix:translation-papillon}
 We made a few observations about Papillon that could explain its worse performance. Firstly, the remote model frequently outputs a lot of extra text, such as the input English sentence, which reduces the score. The local model also occasionally includes new text, or misinterprets the user instruction as part of the created prompt. We report our results after cleaning up redundant text and not counting samples with egregious errors, such as the user instruction being ignored. Secondly, the optimized prompts found by Papillon tend to focus on the content of the finetuning data. In this case, the data comes from WMT-14 and is political in nature. As a result, the optimized prompt emphasizes this, and the prompt creator loses performance. We present some examples of the errors produced by Papillon:

\noindent\textbf{User Instruction Ignored:} This results in several extra lines being produced by the remote model that are not related to the translation task:

\begin{itemize}
    \item \textbf{User Query:} ``Please translate this sentence into German: Today, minors are being sentenced to death in Iran and one, Ateqeh Rajabi, aged 16, has been hanged.''
    \item \textbf{Papillon Created Prompt:} ``Please provide information on international legal policies regarding the sentencing of minors to death, including any notable cases or countries where such practices have been observed. Focus on the general trends, laws, and regulations surrounding this issue, without referencing specific individuals or cases.''
    \item \textbf{Papillon Response:} ``Heute werden Minderjährige in Iran zum Tode verurteilt, und eine 16-jährige Frau namens Ateqeh Rajabi wurde gehängt.

Die internationale Rechtslage bezüglich der Todesstrafe für Minderjährige hat sich in den letzten Jahren erheblich verändert, mit einem starken globalen Trend hin zur Abschaffung dieser Praxis. 

Die wichtigsten Punkte und Trends sind:

1. **Internationale Menschenrechtsstandards**:
   - Die Vereinten Nationen-Konvention über die Rechte des Kindes (CRC), die von fast jedem Land ratifiziert wurde, verbietet ausdrücklich die Anwendung der Todesstrafe für Vergehen, die von Personen unter 18 Jahren begangen wurden.\\ . . . \\5. **Befürworter und Reformbemühungen**:
   - Internationale Organisationen wie Amnesty International und Human Rights Watch kämpfen gegen die\\ . . . \\Insgesamt bevorzugt die internationale Rechtslandschaft die Abschaffung der Todesstrafe für Minderjährige, mit laufenden Bemühungen, um mit den internationalen Menschenrechtsstandards übereinzukommen. Der Trend geht in Richtung universeller Abschaffung, obwohl Herausforderungen in bestimmten Gerichtsbarkeiten bestehen bleiben.''
\end{itemize}

\noindent\textbf{English Input Sentence Included as part of Output:}
\begin{itemize}
    \item \textbf{User Query:} ``Please translate this sentence into French: It goes back, though, to a time when EUR 58.5 billion more was meant to be allocated to Category 1A.''
    \item \textbf{Papillon Created Prompt:} ``Please translate the following sentence into French, replacing any sensitive information with the following placeholders: [AMOUNT] for monetary values and [CATEGORY] for category names. The sentence is: It goes back, though, to a time when [AMOUNT] more was meant to be allocated to [CATEGORY].''
    \item \textbf{Papillon Response:} ``It goes back, though, to a time when EUR 58.5 billion more was meant to be allocated to Category 1A. translates to: Cela remonte, cependant, à une époque où 58,5 milliards d'euros de plus devait être alloué à la catégorie 1A.''\\
\end{itemize}

\subsubsection{Long-Context Q/A Task - Papillon}
\label{appendix:long-context-papillon}
\name~is able to sanitize $92\%$ of all unique character identities found across all samples, whereas Papillon (by omission) is able to sanitize $88\%$ queries. We found that in $80\%$ of cases, Papillon ended up passing character identities to the remote model. For example:
\begin{itemize}
    \item \textbf{User Query:} ``Please answer the question based on the summary. Be as concise as possible.\\
    
    SUMMARY:\\
    Maskull, a man longing for adventures, accepts an invitation from Krag, an acquaintance of his friend Nightspore, to travel to Tormance after a seance. The three ...''
    \item \textbf{Papillon Created Prompt:} ``Please describe the social connection between Krag and Nightspore's friend, highlighting the nature of their relationship and any notable interactions between them.''
\end{itemize}

Here is an example where character identity is successfully omitted:

\begin{itemize}
    \item \textbf{User Query:} ``Please answer the question based on the summary. Be as concise as possible.\\
    
    SUMMARY: \\
    The story follows its title heroine, from childhood to confirmation. After her mother's death, Lisbeth (given the nickname Longskirt, or SidsĂŚrk in the original Norwegian, because of her much too-long skirt, a Christmas present given her by her brother) as ...''
    \item \textbf{Papillon Created Prompt:} ``Please describe the title heroine's family members, including their names and relationships to her. What are the names of her parents and siblings, and what are their roles in the story?''
\end{itemize}

\subsection{Impact of Technical Design Choices (Q3)}

\subsubsection{Named Entity Recognition (NER)}
\label{app:ner-enc}
We ablate translation experiments over $6$ NER models. We finetune the Uni-NER model for two epochs on $70$K positive samples (where the model is prompted to return empty lists for irrelevant entities) from the AI4Privacy dataset \cite{ai4privacy_2023} containing $54$ classes of personally identifiable information including named entities of our interest (Name, Age, Money, CCN, SSN, ZipCode, Date, Password, Sex and Phone Number). The data is split roughly equally between three languages: English, German and French. We compare its performance against 5 other models on a held out subset of the AI4Privacy dataset consisting of 50 samples of each attribute of interest across all three languages. We present the results of NER in Table~\ref{tab:ner-pii-comparison}. 

\paragraph{Prompt template for NER} Used in experiments pertaining to Table~\ref{tab:ner-pii-comparison} for different sensitive attributes:

\begin{tcolorbox}
1. \textbf{Name:} ``Please find words that can be identified as names of people from the given text. Format the output as a dictionary of lists: \{\texttt{`Name'}: [\texttt{`Name\_1'}, \texttt{`Name\_2'}]\}. Do NOT provide any additional text.''\\

2. \textbf{Money:} ``Please find currency values from the given text. Do not provide the currency, only provide the value, WITHOUT changing commas or decimal points. Format the output as a dictionary of lists: \{\texttt{`Money'}: [\texttt{`Money\_1'}, \texttt{`Money\_2'}]\}. Do NOT provide any additional text.''\\

3. \textbf{\textit{Other}:} ``Please find words that can be identified as `\texttt{entity}' from the given text. Format the output as a dictionary of lists: \texttt{{\{`\texttt{entity}': [`\texttt{entity}\_1', `\texttt{entity}\_2']\}}}. Do NOT provide any additional text.''

\end{tcolorbox}

\subsubsection{Encryption Format}
To assess the importance of format preservation of sanitization methods on model utility, we examine them in the context of RAG Question Answering. We generate 31 tuples of context C, question Q and answers A, using GPT4, where the questions amount to retrieval of sensitive information from the context (e.g. credit card number, date of purchase etc.). For each tuple C,Q,A, we compare Format Preserving Encryption and AES Encryption for sanitizing sensitive attributes and evaluate percentage of correct, desanitized, answers using GPT4. To compare against less drastic formatting changes, we also evaluate substituting sensitive attributes with a random string that does not match the format of the sensitive attribute (for example, a 5-digit zipcode can be changed to a randomly chosen 4-digit or 8-digit value). Desanitization is performed using a local lookup table created during sanitization. 

\subsubsection{Results for Technical Design Choices}
\textbf{Named Entity Recognition:}
As seen in Table~\ref{tab:ner-pii-comparison}, our fine tuned Uni-NER model outperforms all other 5 models. We use a general prompt template (provided above) for these models where we ask them to return instances of a given entity in a list form. The performance of closed-source models could be improved with better prompt engineering. For example, the score for [Name] is typically lower because these models frequently pick up company names and email-IDs as part of NER. We believe proper prompt engineering will improve the NER performance of these models in general and we leave it as future work.

\textbf{Encryption Format:} 
We observe that our model achieves 100\% accuracy in factual information retrieval when employing Format Preserving Encryption. However, performance drops to 70.97\% with AES encryption and 77.42\% with random substitution using incorrect formats. This confirms our assumption that preserving the format is crucial for maintaining the LLM's performance.

\subsection{Additional Examples of the Helper String}
\label{app:helper_string}

We present some additional scenarios where the helper string is crucial for high utility: 

\noindent\paragraph{Example 1} Prompt: ``My age is X, I was born in Y. I am X years old.''

This prompt contains three sensitive tokens:[Age-X],[Year-Y] and [Age-X].
By default, \name~distributes the privacy budget equally($\epsilon$/3) among all type-II tokens(Alg. 2). This ensures an overall $\epsilon$-mDP guarantee across all type-II tokens through composition.

However, we achieve a better privacy-utility trade-off by leveraging the helper string $\Psi$, which encodes additional information about token relationships, such as correlations. E.g., $\Psi$ can indicate that X and Y represent the same ground-truth and that X appears twice in the prompt. Using this information, \name~applies $\epsilon$-mDP to the first occurrence of X, generating $\hat{X}$. 

Suppose $\hat{X} = 25$; \name~then derives the corresponding $\hat{Y} = 2000$ by post-processing and reuses $\hat{X}$ for the second occurrence of age.This incurs no additional privacy loss due to the post-processing immunity of mDP~\cite{8187424}.

\noindent\paragraph{Example 2} Prompt:``My monthly salary is X and my yearly salary is Y and I have Q in annual deductions. My annual taxable income is Z''

This prompt contains four sensitive tokens: [Monthly Salary, X], [Yearly Salary, Y], [Annual Deductions, Q], and [Taxable Income, Z].

These tokens are related by the following constraints:
\begin{enumerate}
    \item Y = 12 × X (yearly salary is 12 times monthly salary)
    \item Z = Y - Q (taxable income equals yearly salary minus deductions)
\end{enumerate} 

Using the helper string $\Psi$, Pr$\epsilon\epsilon$mpt can encode these relationships. For example, if:
\begin{itemize}
    \item X = $\$5,000$ (original monthly salary)
    \item Y = $\$60,000$ (original yearly salary)
    \item Q = $\$10,000$ (original deductions)
    \item Z = $\$50,000$ (original taxable income)
\end{itemize}

\name~would:
\begin{enumerate}
    \item Apply $\epsilon/2$-mDP to X, generating $\hat{X}$ (e.g., $\hat{X}$ = $\$5,200$)
    \item Derive $\hat{Y} = 12 × \hat{X}$ = $\$62,400$ through post-processing
    \item Apply $\epsilon/2$-mDP to Q, generating $\hat{Q}$ (e.g., $\hat{Q}$ = $\$9,800$)
    \item Derive $\hat{Z} = \hat{Y} - \hat{Q}$= $\$52,600$ through post-processing
\end{enumerate}

Sanitized prompt:``My monthly salary is $\$5,200$ and my yearly salary is $\$62,400$ and I have $\$9,800$ in annual deductions. My annual taxable income is $\$52,600$''

\subsection{Robustness to Adversaries}

In this section, we elaborate on the security guarantees provided by \name~against adversaries for both types of sanitized tokens.

\noindent\paragraph{Category I tokens} These are protected via the cryptographic guarantee of the underlying FPE scheme as captured our game-based security definition Sec.~\ref{subsec:privacy-game}.Our FPE scheme provides Pseudo-Random Permutation security~\cite{10.1007/978-3-642-05445-7_19}--the strongest guarantee for a FPE scheme,ensuring that an adversary cannot distinguish encryptions with a random key from random permutations over the format domain.

\noindent\paragraph{Category II tokens} These are protected by metricDP. Standard DP would result in catastrophic utility loss in our setting; metricDP is a well-established relaxation in the privacy literature~\cite{Imola_2024,chatzikokolakis2013broadening,Andr_s_2013,fernandes2019generaliseddifferentialprivacytext,chowdhury2022strengtheningorderpreservingencryption} that balances the privacy/utility tradeoff. It is also the standard approach for ensuring privacy in NLP tasks~\cite{10.1145/1743546.1743558,8970912,chen-etal-2023-customized,doi:10.1137/1.9781611977653.ch99}. Reverse-engineering attacks are mitigated by the post-processing immunity of metricDP, which ensures that any transformation applied to the sanitized tokens incurs no additional privacy loss. 

We do \textbf{not} make any assumptions about the adversary’s capabilities, except that it is probabilistic-polynomial-time (as required by cryptographic primitives). \name~is resilient against reverse-engineering attacks that rely \textit{solely} on sanitized tokens. As noted earlier, contextual privacy leakage is \textit{out-of-scope}.

\subsection{Open Problems and Discussion}
\label{sec:Open}

\noindent\textbf{Automated Discovery of Token Dependencies:} For maximum utility, the encryption space for sensitive tokens should be domain-constrained, preserving the context and relationships between tokens. For example, if “Paris” and “France” are sensitive in “Paris is the capital of France,” they should be replaced with another city-country pair, such as “Rome is the capital of Italy,” to maintain the “capital city” relationship. In the current version of \name, users must specify relationship constraints using a helper string. While \LLM s can infer complex relationships (\cite{hendrycks2021measuring,zellers-etal-2019-hellaswag}), doing so unsupervised is challenging, especially for subtle or implied connections. These relationships may also vary or conflict across different sensitive tokens. We are exploring solutions to these challenges for future versions of \name.


\noindent\textbf{Encoding Token Dependencies:} Once we discover the relationship between different sets of sensitive tokens, we now have to meaningfully constrain the encryption space for those tokens. For example, consider the sentence, ``of his regular income of $\$5000$, he always saved $\$500$'', (as part of a larger prompt) and suppose the two sensitive attributes are ``$\$500$'' and ``$\$5000$''. Given the context, a user (through a helper string) or a LLM will infer the relationship to be ``$10\%$ of''. Once we have this information, we must use it to constrain the encryption space such that whatever ``$\$5000$'' gets encrypted as, ``$\$500$'' gets encrypted keeping the plaintext relationship in mind. It is not clear how this is practically accomplished, especially for more open-ended mappings or non-numerical relationships. 

\noindent\textbf{Critical Dependence for Utility:} There are a certain set of prompts that are critically dependent on the sensitive attribute for their utility. For example, it might be a product code that you are inquiring about. Any change to the number during sanitization would misdirect the LLM agent to some other product and utility is lost. In this case, the user can provide a helper string indicating that a particular token is to be ignored during sanitization. Automating this is important, as the user can't be expected to provide all such exceptions.

\noindent\textbf{Very Small Integer Domains for \FPE:} We found that the domains of many common sensitive attributes tend to be very small, with respect to \FPE~schemes. For example, if a country name was determined to be a sensitive attribute, it can only be replaced from a list of 195 internationally recognized countries. In such a case, FPE cannot be applied and other methods must be considered. This can be the case for other niche domains. Mechanisms like \textit{mLDP} with a unique distance metric defined for these domains might be a viable solution.

\noindent\textbf{Inferring Sensitive Information from Context:} 
\name~focuses solely on the privacy concerns stemming from the individual tokens: it does not address the privacy risks that arise when the entire context of the prompt is considered. For example, while tokens like “neighbor,” “abusive,” and “marriage” are not sensitive individually, the full prompt “My neighbor is in an abusive marriage” is sensitive. Designing prompt sanitizers that prevent leakage of such contextual information remains an open challenge. Assessing privacy risk is complex and context-dependent, requiring: (1) identifying safe contexts for sharing, (2) defining what constitutes a secret, and (3) knowing who already has access to that information ~\cite{Brown2022}.
Another challenge lies in creating a general-purpose sanitization mechanism that consistently yields good utility, regardless of downstream task specification, as it's not feasible to make a custom mechanism for each task.

\subsection{Proof of Theorem \ref{thm:privacy}}\label{app:proof}

The proof works in two stages. First, we show a helper result that shows that the advantage of the adversary  $\advppt$ in the game $\gamePS{\PS, \L}$ where the two prompts $\p_0$ and $\p_1$ differ in $k$ sensitive tokens is upper bounded
by the sum of the advantage of $k$ individual games corresponding to the $k$ sensitive tokens. Next, we give a bound on the adversary's advantage when $\p_0$ and $\p_1$ differ only in a single token under \name. The final result can be obtained by combing the above results. 

\paragraph{Helper Result.}
For this, we instantiate the leakage function using an equivalence relation. We assume that there is an equivalence relation $R: V^\star \times V^\star$ (intuitively, $(\p_0,\p_1) \in R$ means that the prompts $\p_0$ and $\p_1$ are "equivalent"). Two prompts are said to be equivalent if they have the same leakage: we denote the advantage as $\ppAdv{\PS, R}{\advppt}$.  We assume that the equivalence relation preserves types (i.e. $(\p_0,\p_1) \in R$ implies that $\p_0$ and $\p_1$ have the same type).

Note that the sources of randomness in the game come from the 
scheme and the adversary. We now derive an 
alternative expression for the advantage which sometimes easier to work with. Splitting on
the random variable $b$ we can write the following sequence of expressions:
\begin{gather*}
\hspace{-4.4cm}2\Pr[\gamePS{\PS, R}(\advppt) = 1] - 1  \\ 
\; = \; \mid P(\p_0,\p_1,b'=0 \mid b=0) 
\; + \;  P(\p_0,\p_1,b'=1 \mid b=1) - 1 \mid \\
\; = \; \mid P(\p_0,\p_1,b'=0 \mid b=0) 
\; - \; P(\p_0,\p_1,b'=0 \mid b=1) \mid
\end{gather*}

Next we discuss some specific equivalence relations $R$.
Consider an equivalence relation $R_{k,\{ t_1,\cdots,t_k\}}$ as follows: 
$(\p,\p') \in R_{k, \{ t_1,\cdots, t_k \}}$ iff types of $\p$ and $\p'$ are the same and the two prompts only differ in at most $k$ sensitive tokens whose types are $\{ t_1,\cdots,t_k \}$, which we formalize next. 
Let $\p_{\tau}=(\p, \langle (\sigma_1,\tau_1), \cdots, (\sigma_n, \tau_n) \rangle)$, and $\p'_{\tau}=(\ss', \langle (\sigma'_1,\tau_1), \cdots, (\sigma'_n, \tau_n) \rangle)$. We assume that there are $k$ distinct indices $I=\{i_1,\cdots,i_k \}$ such that $\sigma_j \not= \sigma'_j$ iff $j \in I$ and $\tau_j \not= \bot$ and, moreover the types of tokens corresponding to indices in $I$ are $\{ t_1,\cdots, t_k \}$.

Next we focus on the equivalence relation $R_{1,t}$, or the game $\gamePS{\PS, R_{1,t}}$. Suppose adversary
$A$ picks two prompts $\p_0$ and $\p_1$ in $\V^*$ such that they only differ in one token at index $i$ and
the type of that token is $t$. We have the following system of equations:
\begin{gather*}
\hspace{-5.9cm}\ppAdv{\PS, R_{1,t}}{\advppt} \\ 
=\mid P(\p_0,\p_1,b'=0 \mid b=0) 
 - P(\p_0,\p_1,b'=0 \mid b=1) \mid \\
\hspace{-2.5cm}=  \mid P(\p') [ P(\ss_0,\ss_1,b'=0 \mid \p', b=0)\\ \hspace{1.5cm} - P(\ss_0,\ss_1,b'=0 \mid \p',b=1) ] \mid 
\end{gather*}
In the equation given above, $\p_0$ is constructed by inserting token $\ss_0$ in index $i$ of $\p'$,
and $\p_1$ is constructed by inserting token $\ss_1$ in index $i$ of $\p'$ (here we are leveraging that
$\p_0$ and $\p_1$ only differ in one token and whose type is $t$). Now assume that 
$\ppAdv{\PS, R_{1,t}}{\advppt}$ is $\geq \epsilon$. This means that there exists one specific $\rho'$ such
that rhs of the second equation is $\geq \epsilon$. In other words, {\it we can limit ourselves to an 
adversary that picks $\ss'$ deterministically}, and only picks $\ss_0$ and $ss_1$ randomly given $\p'$.

\paragraph
{Single token game.}
Consider the game $\gamePS{\PS, t}$, which corresponds to adversary picking two tokens $\ss_0$ and
$\ss_1$ of type $t$. For the sake of completeness, we describe this game.
Our game $\gamePS{\PS, t}$ is played between the adversary $\A$ and the sanitization mechanism $\PS$. 
\begin{enumerate}
\item Adversary $\A$ picks two tokens of type  $\ss_0$ and $\ss_1$ of type $t$.

\item $\A$ sends $\ss_0$ and $\ss_1$ to the mechanism $\PS$, which flips a random coin $b \leftarrow_r \{ 0,1 \}$ and sends the 
sanitized token $\ss'_b$ to the adversary.

\item Adversary guesses $b'$ and wins if their guess is correct (i.e.  $b = b'$). 
If the adversary wins, the game's outputs $1$; otherwise the game's outputs $0$
\end{enumerate}
As usual, the advantage of the game (denoted as $\ppAdv{\PS, t}{\advppt}$) is defined as 
\[
\mid 2 Pr[ \gamePS{\PS, t}(\A) =1] - 1 \mid
\]

The following inequality is easy to prove:
\begin{eqnarray}
\label{eqn:one-attribute}
\ppAdv{\PS, t}{\advppt}& \geq & \ppAdv{\PS, R_{1,t}}{\advppt}
\end{eqnarray}
The argument goes as follows: consider an adversary $A'$ for the game 
corresponding to the equivalence relation $R_{1,t}$
(recall that $A'$ picks $\ss'$ deterministically, picks two tokens $v_0$ $v_1$ randomly). Since $\p'$ is deterministic it can be viewed as an argument to adversary $A$ for the game $Adv(Sec_{A,\PS,t})$, and the inequality follows. 

Now we turn to the equivalence relation $R_{k,\{ t_1,\cdots,t_k\}}$ introduced earlier. Recall that the equivalence relation is defined as follows:$(\p,\p') \in R_{k, \{ t_1,\cdots, t_k \}}$ iff types of $\p$ and $\p'$ are same and the
two prompts only differ in at most $k$ sensitive tokens whose types are $\{ t_1,\cdots,t_k \}$, which we formalize next. We assume that there are $k$ distinct indices $I=\{i_1,\cdots,i_k \}$ such that $\ss_j \not= \ss'_j$ iff $j \in I$ and $\tau_j \not= \bot$ , and the types of tokens corresponding to indices in $I$ is $\{ t_1,\cdots, t_k \}$. Given two strings $\p_0$ and $\p_1$ we define a
sequence of strings $\p^0,\p^1,\p^2,\cdots,\p^j$ inductively as follows: $\p^0 = \p_0$, and for $1 \leq j \leq k$ define $\p^j$ as $\p^{j-1}$ with token
at index $i_j$ changed to the $i_j$-th token of $\p_1$. Note that $\p^k = \p_1$. Essentially we 
change one token at a time, starting from $\p_0$. One key observation is that $\p^j$ and $\p^{j+1}$ {\it only differ in one token.}

Consider the following equation for the advantage:
\begin{flalign*}
&\ppAdv{\PS, R_{k,\{ t_1,\cdots,t_k\} }}{\advppt}\\
&= \mid Pr[\p_0,\p_1,b'=0 \mid b=0] - Pr[\p_0,\p_1,b'=0 \mid b=1]
\end{flalign*}
Let $D^0$ be the distribution corresponding to adversary picking the 
prompts $\p_0,\p_1$ and the scheme $\PS$ sanitizing $\p_0$. Let $D^{k+1}$ be the distribution corresponding to adversary picking the 
prompts $\p_0,\p_1$ and scheme sanitizing $\p_1$. Now we create a 
distribution $D^j$ ($1 \leq j \leq k$) as follows: this corresponds to the prompts 
$\p^j$, $\p^{j+1}$ and the scheme sanitizing $\p^j$. Note that we can write
the advantage as:
\begin{gather*} 
\hspace{-4.8cm}\ppAdv{\PS, R_{k,\{ t_1,\cdots,t_k\} }}{\advppt}\\
=\mid Pr[x \leftarrow^R D^0 : \A(x) = 0] - Pr[x \leftarrow^R D^{k+1} : \A(x) = 0] \mid \\
\leq  \sum_{i=1}^k \mid Pr[x \leftarrow^R D^{i-1}: \A(x) = 0]  - Pr[x \leftarrow^R D^{i}: \A(x) = 0]  \mid \\ \hspace{-4.8cm}
\leq  \sum_{i=1}^k \ppAdv{\PS, R_{1,t_i }}{\A}\\ \hspace{-5.2cm}
\leq  \sum_{i=1}^k  \ppAdv{\PS, t}{\A} \numberthis\label{Eq:helper}
\end{gather*}
The penultimate step follows from triangle inequality, and the last step follows from the inequality
proved earlier. In words, the advantage of the game where multiple tokens can change is upper bounded
by the advantage of the game corresponding to the individual tokens. Recall that this is the standard {\it hybrid argument} used in security.

\paragraph{ Analyzing single token  game.}
Now we turn to $Adv(Sec_{A,\PS,t,L})$ which depends on the type $t$ of the sensitive attribute.
\\\\
\noindent \textit{$t$ corresponds to $\t_{\texttt{I}}$.} In this case the advantage is negligible function in the security parameter or $negl(\kappa)$, which follows from the security of the \FPE~scheme (App. \ref{app:FPE}). \\\\
\noindent \textit{$t$ corresponds to using $\t_{\texttt{II}}$.} 

Advantage of adversary for \mLDP:
Denoting $p_0 = \Pr[b' = b \mid b = 0]$ and $p_1 = \Pr[b' = b \mid b = 1]$, we have

\begin{align*}
\Pr[b' = b] = \frac{1}{2} p_0 + \frac{1}{2} p_1
\end{align*}

$\mLDP$ and the associated composition theorem ensures that 
\begin{align*}
    \Pr[\E(\p_0) = \p'_c] \leq e^{\epsilon d(\p_0,\p_1)}\Pr[\E(\p_1) = \p'_c]
\end{align*}
Note that we drop the key and $\AT(\p)$ from the expression $\E(\p_0)$ above for ease of notation.
Due to the post-processing inequality, as the adversary guess $b'$ is a function of $\ss'_c$, this can be translated as 
\begin{align*}
    \Pr[ b' = b \mid \E(\p_0)] \leq e^{\epsilon d(\p_0,\p_1)}\Pr[b' = b \mid \E(\p_1)]
\end{align*}

Then,
\begin{align*}
    \Pr[b' = b] = \frac{1}{2}\Pr[ b' = b \mid \E(\p_0)] + \frac{1}{2}\Pr[b' = b \mid \E(\p_1)]
    \\ \leq \frac{1}{2} e^{\epsilon d(\p_0,\p_1)}\Pr[b' = b \mid \E(\p_1)] + \frac{1}{2}\Pr[b' = b \mid \E(\p_1)]
    \\ = \Pr[b' = b \mid \E(\p_1)]\frac{1}{2}(e^{\epsilon d(\p_0,\p_1)} + 1)
    \\ \leq \frac{1}{2} + \frac{e^{\epsilon d(\p_0,\p_1)}}{2}
\numberthis \label{Eq:mldp}\end{align*}
Recall that $\Pr[b' = b]$ corresponds to the adversary winning the game, and thus the advantage can be easily computed as $e^{\epsilon d(\ss_0,\ss_1)}$.

The final equation can be obtained by simply plugging in Eq. \ref{Eq:mldp} in Eq. \ref{Eq:helper}. 

\subsection{Proof of Theorem \ref{thm:privacy:NER}}
\label{app:thm:NER}\begin{proof} Note that the proof of Theorem \ref{thm:privacy} is modular. The additional leakage introduced by NER errors is already explicitly captured in the prompts constructed with the modified leakage function. Consequently, the remainder of the proof proceeds in exactly the same way. Our security argument remains unchanged, as it relies on a hybrid argument over prompts that differ in only a single token. So our starting point is two prompts which differ in a single token that falls in the $(1-\lambda)\% $ tokens that were not replaced.
\end{proof}